\documentclass[english]{iopart}
\pdfoutput=1
\usepackage[T1]{fontenc}
\usepackage[utf8]{inputenc}
\usepackage{babel}
\usepackage{verbatim}
\usepackage{float}
\usepackage{amsmath}
\usepackage{amssymb}
\usepackage{bm}
\usepackage{graphicx}
\usepackage{booktabs}
\usepackage{graphicx}
\usepackage{subscript}
\usepackage[unicode=true,pdfusetitle,
bookmarks=true,bookmarksnumbered=false,bookmarksopen=false,
breaklinks=false,pdfborder={0 0 1},backref=false,colorlinks=false]
{hyperref}
\usepackage{xcolor}
\usepackage[normalem]{ulem}

\makeatletter

\newcommand{\lyxmathsym}[1]{\ifmmode\begingroup\def\b@ld{bold}
	\text{\ifx\math@version\b@ld\bfseries\fi#1}\endgroup\else#1\fi}

\usepackage{geometry} 
\usepackage{xparse}

\newcommand{\ket}[1]{\ensuremath{\left|#1\right\rangle}}

\newcommand{\matrixel}[3]{\ensuremath{\left\langle #1 \vphantom{#2#3} \middle| #2 \middle| #3 \vphantom{#1#2} \right\rangle}}

\newmuskip\pFqmuskip
\newcommand*\pFq[6][8]{%
	\begingroup 
	\pFqmuskip=#1mu\relax
	\mathchardef\normalcomma=\mathcode`,
	\mathcode`\,=\string"8000
	\begingroup\lccode`\~=`\,
	\lowercase{\endgroup\let~}\pFqcomma
	{}_{#2}F_{#3}{\left[\genfrac..{0pt}{}{#4}{#5};#6\right]}%
	\endgroup
}
\newcommand{\pFqcomma}{{\normalcomma}\mskip\pFqmuskip}

\ExplSyntaxOn
\NewDocumentCommand{\MeijerG}{smmmm}
{
	\IfBooleanTF{#1}
	{
		\vic_meijerg:nnnnnn { #2 } { #3 } { #4 } { #5 } { small } { }
	}
	{
		\vic_meijerg:nnnnnn { #2 } { #3 } { #4 } { #5 } { } { \; }
	}
}

\seq_new:N \l__vic_meijerg_args_in_seq
\seq_new:N \l__vic_meijerg_args_out_seq

\cs_new_protected:Nn \vic_meijerg:nnnnnn
{
	\seq_set_split:Nnn \l__vic_meijerg_args_in_seq { | } { #3 }
	\seq_clear:N \l__vic_meijerg_args_out_seq  
	\seq_map_inline:Nn \l__vic_meijerg_args_in_seq
	{
		\seq_put_right:Nn \l__vic_meijerg_args_out_seq
		{
			\begin{#5matrix} ##1 \end{#5matrix}
		}
	}
	G\sp{#1}\sb{#2}
	\left(
	\seq_use:Nn \l__vic_meijerg_args_out_seq { #6\middle|#6 }
	#6\middle|#6
	#4
	\right)
}
\ExplSyntaxOff

\makeatother

\begin{document}
\title{Theoretical methods for excitonic physics in two-dimensional materials}

\author{M. F. C. Martins Quintela$^{1,2}$,  J. C. G. Henriques$^{1}$,   and N. M. R. Peres$^{1,2}$}
\address{$^{1}$Department and Centre of Physics, University of Minho, Campus of Gualtar, 4710-057, Braga, Portugal}
\address{$^{2}$International Iberian Nanotechnology Laboratory (INL), Av. Mestre Jos{\'e} Veiga, 4715-330, Braga, Portugal}
\begin{abstract}
	In this tutorial we introduce the reader to several theoretical
	methods of determining the exciton wave functions and the corresponding
	eigenenergies. The methods covered are either analytical, semi-analytical,
	or numeric. We make explicit all the details associated with the different
	methods,   thus allowing newcomers to do research on their own, without experiencing a steep learning curve. The tutorial starts with a variational
	method and ends with a simple semi-analytical approach to solve the
	Bethe-Salpeter equation in two-dimensional (2D) gapped materials. For
	the first methods addressed in this tutorial, we focus on a single layer of
	hexagonal Boron Nitride (hBN) and of transition metal dichalcogenide (TMD), as these are exemplary materials in the field of 2D excitons.  For explaining the Bethe-
	Salpeter method we choose the biased bilayer graphene, which presents
	a tunnable band gap. The system has the right amount of complexity
	(without being excessive).  This allows the  presentation of the solution of the Bethe-Salpeter equation (BSE) in a context that can be easily generalized to more
	complex systems or to apply it to simpler models.
\end{abstract}


\section{Introduction}

The dawn of two-dimensional (2D) materials occurred with the isolation of a single graphene layer on a silicon oxide substrate \cite{doi:10.1126/science.1102896}. Around the same time, other 2D materials were isolated. Among these we find hexagonal boron nitride (hBN)\cite{doi:10.1126/science.1091979}, transition metals dichalcogenides (TMD)\cite{RevModPhys.90.021001}, and 2D superconducting layers \cite{2DSup}. 

The excitement about the new physics found in graphene layers, such as the half-integer quantum Hall effect (also found at room temperature) \cite{JIANG200714,Fujita2016}, Klein tunnelling \cite{PhysRevLett.102.026807}, ballistic electronic transport \cite{Baringhaus2014}, high thermal conductivity \cite{doi:10.1063/1.2907977}, constant optical absorption in the visible range of the electromagnetic spectrum \cite{fine_visual,doi:10.1063/1.3073717}, and highly confined graphene plasmons with long propagation length \cite{doi:10.1021/nl201771h,Legrand:17,Zhang2022,Ni2018}, shadowed for some time the interest in other 2D materials. As graphene physics matured, the interest in other 2D materials grew and  novel physical properties were found.  The existence of correlated phases in twisted graphene layers \cite{Liao_2021,Choi2021Nature,Choi2021NatPhys,Yu2022,doi:10.1126/science.aav1910,Cao2020}, valley physics and excitons in TMDs \cite{Wang_2016,Maragkou_2015,Wang_2018,Krenner_2005,RevModPhys.90.021001,Wang2021,PhysRevLett.127.106801}, 2D superconductivity \cite{PhysRevB.104.035104,PhysRevLett.127.247001,cea2021superconductivity,chatterjee2021intervalley,Zhou2021,cea2021superconductivity},  2D magnetic materials, and the myriad of heterostructures made of these materials, has created a new and exciting research field in condensed matter physics and materials science alike.  

Among the several correlated phases in 2D materials, excitons are of particular interest for a new class of opto-electronic devices.  In the simplest possible picture, an exciton corresponds to a bound electron-hole pair which is formed when an electron is promoted to the conduction band, thus leaving a hole in the valence band. These two \emph{particles}, having opposite charges, interact via an electrostatic potential and may create bound states, similarly to what one finds for the Hydrogen atom due to the proton-electron interaction. Detailed experiments \cite{Wang2020,doi:10.1126/sciadv.abg0192,Wei2016} revealed that the excitonic optical spectrum (characterized, for example, by how the systems absorbs light) is rather different from what is predicted using the 2D Schrödinger equation with an attractive Coulomb potential \cite{PhysRevB.84.075439,PhysRevB.103.045124,PhysRevLett.111.216805}.   Due to the reduced thickness of 2D materials, the electrostatic interaction between the electron and the hole is far less screened, and therefore stronger, than what is found for 3D systems. This allows for the emergence of stable excitons, which may be observed even at room temperature \cite{https://doi.org/10.1002/adom.202101305,doi:10.1021/nl503799t,Calman2018,Shan2021,Mueller2018} (something harder to realize in conventional semiconductors). Besides excitons, other excitations may appear resulting from the interaction of one or more electrons, with one or more holes. The most prominent example of these, are the trions observed in doped TMDs, resulting from the interaction of two electron with one hole \cite{KUECHLE2021100097,Li2018}. 

The scope of correlated phase in 2D materials is currently so vast that it is difficult to cover it within a single paper. Therefore, in this tutorial, we choose to focus our attention on the theoretical methods necessary to address the calculation of the properties of excitons in gapped 2D materials, with special emphasis on hBN, TMDs, and biased graphene bilayer. The latter example will be tackled using the Bethe-Salpeter equation proving a good example of a multiband system with a moderate degree of complexity. This will allow the reader to embark both on more complex systems (such as the biased ABC trilayer, with six bands) or on simpler systems with less bands. For every presented approach, our starting point is the tight-binding formulation for the single particle spectrum, thus avoiding the need of using {\it ab-initio} methods.

\section{Optical properties of gapped 2D materials}

\begin{figure*}
	\centering{}\includegraphics[scale=1.3]{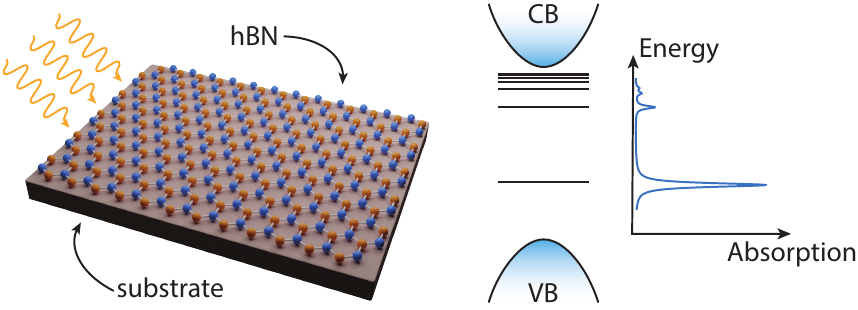}
	\caption{\label{fig:hbn_exciton_panel}\textbf{Excitonic resonances in hBN bandgap.} (Left) Top view of a hBN monolayer deposited on a substrate illuminated by electromagnetic radiation. (Right) Diagramatic view of the various bound electron--hole pairs in the bandgap, together with the various absorption peaks characteristic of each excitonic resonance. In the diagram, VB stands for valence band while CB stands for conduction band. 
	}
\end{figure*}

Monolayer hBN is a 2D material with a honeycomb lattice, known for being an  electronic insulator with a band gap of approximately 7 eV \cite{Rom_n_2021}, and whose electronic properties are essentially determined by a single pair of bands. The large band gap, and the simplicity of the electronic structure makes hBN an ideal system to explore excitonic effects. In Fig. \ref{fig:hbn_exciton_panel} we schematically show the bands of hBN and the excitonic bound states that appear inside the band gap. These states become increasingly close to each other as one reaches the bottom of the conduction band, and are responsible for the optical activity inside the gap.

In a similar fashion to monolayer hBN, the family of monolayer TMDs is a prototypical example of systems where the excitonic effect can be studied. These semiconducting systems present a well resolved band gap of  hundreds of  meV, and their band structure can still be treated with a two band model, if one is concerned with a single spin orientation. Contrary to hBN, TMDs present a strong spin-orbit coupling effect, leading to distinct energy dispersions at the two non-equivalent vertices of its Brillouin zone \cite{RevModPhys.90.021001}. This effect is responsible for the so called valley-physics, which gave rise to the field of valleytronics  \cite{RevModPhys.90.021001,Schaibley2016}.

Despite its interesting features, single layer graphene is a gapless material and, therefore, hosts no bound electron-hole pairs. Nonetheless, the interaction of unbound electrons and holes still introduces modifications in high energy part of the optical response \cite{PhysRevLett.112.207401}, the most noticeable being the change in energy of the van Hove singularity.

In stark contrast with its monolayer counterpart, biased bilayer graphene shows a tunable gap which can be as large as ~150 meV \cite{doi:10.1021/nl902932k}. The magnitude of this band gap can be controlled via an external electric field and dielectric environment,  and allows the formation of tunable excitons, which have already been measured both experimentally \cite{doi:10.1126/science.aam9175} and described theoretically \cite{PhysRevB.105.045411}.  Contrarily to monolayer hBN and monolayer TMDs, the optical response of biased bilayer graphene is dominated by four bands, instead of just two. Despite being more intricate to describe theoretical, this type of multiband system is still tractable with simple methods.

As noted monolayer hBN  presents a band gap that allows the formation of bound electron-hole pairs in the gap,  this is one of the simplest systems where excitons can be studied.  The formation of 
energy levels in hBN is represented  in a simplified manner in Fig. \ref{fig:hbn_exciton_panel}, together with schematic optical absorption spectrum inside the gap, clearly showcasing each excitonic resonance. These resonances become increasingly close as one reaches the top of the bandgap, as can be seen in the right panel of Fig. \ref{fig:hbn_exciton_panel}.

Twisted graphene bilayer presents no gap under the application of an external electric field perpendicular to the layers, but its band structure can still be tuned by controlling the twist angle as well as the magnitude of the electric field \cite{PhysRevLett.99.256802,Li2010,PhysRevB.86.155449,Andrei2020,ZHOU2022115204}. On the same trend, Bernal stacking trilayer graphene (also known as ABA trilayer graphene) does not show a gap in the same conditions, with the electric field only modifying the effective mass of the electrons. Interestingly, rhombohedral  trilayer graphene (also known as ABC trilayer graphene) does open a gap when an electric field is applied to the system \cite{Lui2011}. Therefore, this system is able to sustain bound electron-hole pairs, which, as in biased bilayer graphene are tunable through the magnitude of the electric field. 

In this tutorial we will guide the reader through the determination of the excitonic properties in simple three band systems. These are: monolayer hBN and monolayer TMDs, as well as for biased bilayer graphene. The latter already shows the complexity of a multiband system without being exceedingly difficult to treat. The ABC trilayer graphene is left as an exercise for the adventurous reader.

\section{General formalism}
In this section we shall discuss the general equations that define the excitonic problem in 2D materials.

\subsection{The first approach \label{subsec_Wannier}}
The Wannier equation was first introduced in 1937 by Wannier 
\cite{Wannier1937}s
 to describe {\it the structure of electronic excitation levels in insulating crystals}. It is, essentially, a Schrödinger equation for a 3D hydrogen atom with an effective reduced mass. Such an approach clearly tries to capture the hydrogen-like nature of the electron-hole interaction in solids. Although first introduced for the 3D systems, it can easily be expressed for 2D systems. Generically the Hamiltonian in such a description reads
\begin{equation}
	H=-\frac{\hbar^2\nabla^2}{2\mu_{eh}}-V(\mathbf{r})\,,
	\label{Sch_eq}
\end{equation}
where $\mu_{eh}$ is the reduced mass of the electron-hole pair and $V(\mathbf{r})$ is the electrostatic potential among electrons. The goal is to determine the wave function and eigenenergies of the equation $H\psi(\mathbf{r})=E\psi(\mathbf{r})$, where $\mathbf{r}$ is the relative position vector of the electron and hole, $E$ the exciton energy and $\psi(\mathbf{r})$ its wave function. When $V(\mathbf{r})$ is the attractive Coulomb potential, the energy levels read in 3D
\begin{equation}
	E_n=-\frac{\mu_{eh}c^2\alpha^2}{2\epsilon_r^2n^2},
	\label{eq:hydrogen_levels}
\end{equation}
and \cite{PhysRevA.43.1186,portnoi}
\begin{equation}
	E_n=-\frac{\mu_{eh}c^2\alpha^2}{2\epsilon_r^2(n-1/2)^2},
	\label{eq:hydrogen_levels_2D}
\end{equation}
in 2D,
with $n=1,2,3,\dots$ the principal quantum number, $\alpha\approx 1/137$ is the fine structure constant,  $c$ the speed of light,  and $\epsilon_r$ the relative permittivity of the semiconductor crystal (in 2D,  $\epsilon_r$ is the environment dielectric constant).  Although this model gives good results for the 3D case, its description of 2D system is not ideal, since excitons in these materials do not follow a Rydberg series as Eq. (\ref{eq:hydrogen_levels_2D}) would indicate. Despite its flaws, this direct application of the Wannier model to 2D systems already suggests that excitons should be more tightly bound in 2D than those found in 3D crystals. While for conventional crystals, excitons present binding energies of just a few tens o meV \cite{excitons_3D} (and are subject to thermal dissociation at room temperature), excitons in 2D materials are far more robust.

When transformed to momentum space, Eq. (\ref{Sch_eq}) acquires the form

\begin{equation}
	\frac{\hbar^2 k^2}{2\mu_{eh}}\phi(\mathbf{k})-\sum_{\mathbf{k}^\prime}V_{\mathbf{k}-\mathbf{k}^\prime}\phi(\mathbf{k}^\prime)=E_n\phi(\mathbf{k})\label{eq:Wannier_Momentum}
\end{equation}
As we shall see below, this equation is rather similar to the Bethe-Salpter equation (BSE) which accurately describes the excitonic properties in 2D materials, apart from a missing multiplicative term on the left hand side. Here, $\phi(\mathbf{k})$ and $V_{\mathbf{k}-\mathbf{k}^\prime}$ are the Fourier transforms of $\psi(\mathbf{r})$ and $V(\mathbf{r})$, respectively, where $V(\mathbf{r})$ is the repulsive electron-electron interaction.

\subsection{Bethe-Salpeter equation}
In this part of the text we shall give the main steps to derive the Bethe-Salpeter equation for a simple two band model. In the end we shall see how this equation compares with the Wannier model given above, and how it extends to the multiband case.

We start  defining the exciton creation operator in a two band system as
\begin{equation}
	b_{\nu}^{\dagger}=\frac{1}{\sqrt{A}}\sum_{\mathbf{k}}\phi_{\nu}(\mathbf{k})c_{\mathbf{k},c}^{\dagger}c_{\mathbf{k},v},
	\label{eq:exciton_operator}
\end{equation}
where $A$ is the area of the monolayer, $\phi_{\nu}(\mathbf{k})$
is the Fourier transform of the exciton wave function (unknown at this point) whose quantum numbers we label
by $\nu$.  For isotropic systems, and inspired by the problem of the Hydrogen atom, is customary to separate the exciton wave function into a radial and an angular part, as $\phi_\nu(\mathbf{k})=f_\nu(k) e^{im\theta}$, where $m=0,\pm1,\pm2,...$ stands for the angular quantum number of the exciton \cite{Chaves_2017}. A similar type of solution is found when dealing with the real space wave function. The operator $c_{\mathbf{k},c}^{\dagger}/c_{\mathbf{k},v}$
creates/annihilates an electron in the conduction band with momentum
$\mathbf{k}$.  Since $b_{\nu}^{\dagger}$ is defined in terms of the product
of two electron operators, it obeys bosonic commutation relations;
excitons are, at least approximately, bosons.   The reasoning behind this definition for the exciton operator is simple. As previously mentioned, an exciton is formed when an electron is promoted from the valence to the conduction band, or, in other words, an electron is annihilated in the valence band, and one is created directly above it on the conduction band (we assume  only vertical transitions due to the small momentum carried by photons). In principle, this transition between bands can occur in different points of the reciprocal space. Thus, to be general, we can express the exciton creation operator as a superposition of $c_{\mathbf{k},c}^{\dagger} c_{\mathbf{k},v}$ in momentum space modulated by a given function, in this case $\phi_{\nu}(\mathbf{k})$, which is yet to be determined.

When this operator acts
on the excitonic ground state (or excitonic vacuum), which corresponds to the state with a filled valence band and an empty conduction band (no excitons present), we find
\begin{equation}
	b_{\nu}^{\dagger}|GS\rangle=|\nu\rangle
\end{equation}
where $|\nu\rangle$ represents the state of the exciton we have just
created. We write the energy of such a state as $E_{\nu}$ .

To obtain the equation that gives both $E_\nu$ and $\phi_\nu (\mathbf{k})$, we proceed in the following manner: i) first, we introduce the fermionic Hamiltonian, $H$, of a system of interacting electrons written in second quantization; ii) then, we state that this Hamiltonian should be diagonal when expressed in terms of $b_\nu$; iii) afterwards, we compute the commutator $[H,b^\dagger_\nu$ using the fermionic and bosonic definitions of the two operators; iv) finally, we demand both results to be the same, and obtain the BSE.

We shall now follow the steps given above in a more detailed manner. The electronic Hamiltonian is composed of a kinetic and a potential
contribution. The kinetic term can be readily quantized in terms of
electron operators and reads
\begin{equation}
	\hat{H}_{0}=\sum_{\mathbf{k},\lambda}E_{\mathbf{k},\lambda}c_{\mathbf{k},\lambda}^{\dagger}c_{\mathbf{k},\lambda},
\end{equation}
where the sum over $\lambda$ runs over the bands of the system, and
$E_{\mathbf{k},\lambda}$ is the energy of an electron belonging to
the band $\lambda$, with momentum $\mathbf{k}$ . The potential energy term (two-body interaction)
reads
\begin{align}
	\hat{H}_{{\rm int}} & =\frac{1}{2A}\sum_{\lambda}\sum_{\mathbf{q},\mathbf{k}_{3},\mathbf{k}_{4}}V_{\mathbf{q}}F_{\lambda_{1},\lambda_{2},\lambda_{3},\lambda_{4}}(\mathbf{k}_{3},\mathbf{k}_{4},\mathbf{q})\nonumber \\
	& \times c_{\mathbf{k}_{4}+\mathbf{q}\lambda_{1}}^{\dagger}c_{\mathbf{k}_{3}-\mathbf{q}\lambda_{2}}^{\dagger}c_{\mathbf{k}_{3}\lambda_{3}}c_{\mathbf{k}_{4}\lambda_{4}}
\end{align}
where $V_{\mathbf{q}}$ is the Fourier transform of the interaction
potential among the electrons and
\begin{align*}
	F_{\lambda_{1},\lambda_{2},\lambda_{3},\lambda_{4}}(\mathbf{k}_{3},\mathbf{k}_{4},\mathbf{q}) & =u_{\mathbf{k}_{4}+\mathbf{q}\lambda_{1}}^{\dagger}u_{\mathbf{k}_{3}-\mathbf{q}\lambda_{2}}^{\dagger}u_{\mathbf{k}_{3}\lambda_{3}}u_{\mathbf{k}_{4}\lambda_{4}}\\
	& =u_{\mathbf{k}_{3}-\mathbf{q}\lambda_{2}}^{\dagger}u_{\mathbf{k}_{3}\lambda_{3}}u_{\mathbf{k}_{4}+\mathbf{q}\lambda_{1}}^{\dagger}u_{\mathbf{k}_{4}\lambda_{4}}
\end{align*}
is a form factor with $u_{\mathbf{k}\lambda}$ the $\lambda-$band
spinor, evaluated at $\mathbf{k}$, which is obtained from the eigenvectors
of the $\mathbf{k}$-space model Hamiltonian (see discussion of the biased bilayer further ahead).  We exclude the term
$\mathbf{q}=0$ from the summation the since it cancels with the background contribution due to ion-ion and electron-ion interaction in the solid.

To obtain the equation that defines the exciton wave functions we
will assume that the total Hamiltonian, $\hat{H}=\hat{H}_{0}+\hat{H}_{{\rm int}}$,
is diagonal when written in terms of the exciton operators $b_{\nu}^{\dagger}$, that is, $\hat{H}=\sum_{\nu}E_{\nu}b_{\nu}^{\dagger}b_{\nu}$. Then we will compute
the commutator $[\hat{H},b_{\nu}^{\dagger}]$ using the excitonic
representation of the operator, as well as using their fermionic
representation,  Eq.  (\ref{eq:exciton_operator}). In the end we demand both results to be equal, and
by doing so, an equation defining the exciton wave function emerges.

The computation of $\left[\hat{H},b_{\nu}^{\dagger}\right]$ using
the bosonic representation of the Hamiltonian is trivial, and its
result reads:
\begin{align}
	\left[\hat{H},b_{\nu}^{\dagger}\right] & =E_{\nu}b_{\nu}^{\dagger}\nonumber \\
	& =E_{\nu}\frac{1}{\sqrt{A}}\sum_{\mathbf{k}}\phi_{\nu}(\mathbf{k})c_{\mathbf{k},c}^{\dagger}c_{\mathbf{k},v}.
\end{align}

Contrarily to this commutator, computing $\left[\hat{H},b_{\nu}^{\dagger}\right]$
using the fermionic representations of $\hat{H}$ and $b_{\nu}^{\dagger}$
is a rather cumbersome process. In order to compute this, we will
break the commutator in two, separating contributions those from $\hat{H}_{0}$
and $\hat{H}_{{\rm int}}$. Starting with the evaluation of $\left[\hat{H_{0}},b_{\nu}^{\dagger}\right]$,
and using the following relations regarding commutators and anticommutators
\begin{align}
	[AB,C]= & A[B,C]+[A,C]B,\label{eq:Comm_rel_1}\\{}
	[AB,C]= & A\{B,C\}-\{A,C\}B,\label{eq:Comm_rel_2}
\end{align}
one finds:

\begin{align}
	\left[\hat{H}_{0},b_{\nu}^{\dagger}\right] & =\sum_{\mathbf{k}}\frac{\phi_{\nu}(\mathbf{k})}{\sqrt{A}}\left(E_{\mathbf{k},c}-E_{\mathbf{k},v}\right)c_{\mathbf{k},c}^{\dagger}c_{\mathbf{k},v}
\end{align}
To compute $\left[\hat{H}_{{\rm int}},b_{\nu}^{\dagger}\right]$,
we start  writing
	\begin{align}
		\left[\hat{H}_{{\rm int}},b_{\nu}^{\dagger}\right]=  \sum_{\lambda}\sum_{\mathbf{q},\mathbf{k}_{3},\mathbf{k}_{4},\mathbf{k}}\frac{\phi_{\nu}(\mathbf{k})}{2A^{3/2}}V_{\mathbf{q}}F_{\lambda_{1},\lambda_{2},\lambda_{3},\lambda_{4}}(\mathbf{k}_{3},\mathbf{k}_{4},\mathbf{q})
		\left[c_{\mathbf{k}_{4}+\mathbf{q}\lambda_{1}}^{\dagger}c_{\mathbf{k}_{3}-\mathbf{q}\lambda_{2}}^{\dagger}c_{\mathbf{k}_{3}\lambda_{3}}c_{\mathbf{k}_{4}\lambda_{4}},c_{\mathbf{k},c}^{\dagger}c_{\mathbf{k},v}\right].
	\end{align}
Using Eqs. (\ref{eq:Comm_rel_1}) and (\ref{eq:Comm_rel_2}), it is
possible to show that the commutator can be transformed into:
\begin{align*}
	& \left[c_{\mathbf{k}_{4}+\mathbf{q}\lambda_{1}}^{\dagger}c_{\mathbf{k}_{3}-\mathbf{q}\lambda_{2}}^{\dagger}c_{\mathbf{k}_{3}\lambda_{3}}c_{\mathbf{k}_{4}\lambda_{4}},c_{\mathbf{k},c}^{\dagger}c_{\mathbf{k},v}\right]\\
	= & c_{\mathbf{k}_{4}+\mathbf{q}\lambda_{1}}^{\dagger}c_{\mathbf{k}_{3}-\mathbf{q}\lambda_{2}}^{\dagger}c_{\mathbf{k}_{3}\lambda_{3}}c_{\mathbf{k},v}\delta_{\mathbf{k}_{4},\mathbf{k}}\delta_{\lambda_{4},c}\\
	- & c_{\mathbf{k}_{4}+\mathbf{q}\lambda_{1}}^{\dagger}c_{\mathbf{k}_{3}-\mathbf{q}\lambda_{2}}^{\dagger}c_{\mathbf{k}_{4}\lambda_{4}}c_{\mathbf{k},v}\delta_{\mathbf{k}_{3},\mathbf{k}}\delta_{\lambda_{3},c}\\
	+ & c_{\mathbf{k},c}^{\dagger}c_{\mathbf{k}_{4}+\mathbf{q}\lambda_{1}}^{\dagger}c_{\mathbf{k}_{3}\lambda_{3}}c_{\mathbf{k}_{4}\lambda_{4}}\delta_{\mathbf{k}_{3}-\mathbf{q},\mathbf{k}}\delta_{\lambda_{2},v}\\
	- & c_{\mathbf{k},c}^{\dagger}c_{\mathbf{k}_{3}-\mathbf{q}\lambda_{2}}^{\dagger}c_{\mathbf{k}_{3}\lambda_{3}}^ {}c_{\mathbf{k}_{4}\lambda_{4}}\delta_{\mathbf{k}_{4}+\mathbf{q},\mathbf{k}}\delta_{\lambda_{1},v}.
\end{align*}
where the Kronecker-$\delta$'s appear as a consequence of the anticommutation
relation that the fermionic operators must obey. At this point in
the calculation we are faced with four terms containing the product
of four electron operators. Dealing with such terms is no easy task,
and to simplify the calculation we introduce a mean field approximation.
This technique essentially allows us to transform the previous equation
into one where only products of two operators appear, by taking expectation
values over pairs of operators. For clarity we show how this mean
field approach is applied to first term of the previous equation,
and note that the other terms can be treated in an analogous manner.
We write:
\begin{align*}
	& c_{\mathbf{k}_{4}+\mathbf{q}\lambda_{1}}^{\dagger}c_{\mathbf{k}_{3}-\mathbf{q}\lambda_{2}}^{\dagger}c_{\mathbf{k}_{3}\lambda_{3}}c_{\mathbf{k},v}\\
	\approx & \left\langle c_{\mathbf{k}_{4}+\mathbf{q}\lambda_{1}}^{\dagger}c_{\mathbf{k},v}\right\rangle c_{\mathbf{k}_{3}-\mathbf{q}\lambda_{2}}^{\dagger}c_{\mathbf{k}_{3}\lambda_{3}}\\
	+ & c_{\mathbf{k}_{4}+\mathbf{q}\lambda_{1}}^{\dagger}c_{\mathbf{k},v}\left\langle c_{\mathbf{k}_{3}-\mathbf{q}\lambda_{2}}^{\dagger}c_{\mathbf{k}_{3}\lambda_{3}}\right\rangle \\
	- & \left\langle c_{\mathbf{k}_{4}+\mathbf{q}\lambda_{1}}^{\dagger}c_{\mathbf{k}_{3}\lambda_{3}}\right\rangle c_{\mathbf{k}_{3}-\mathbf{q}\lambda_{2}}^{\dagger}c_{\mathbf{k},v}\\
	- & c_{\mathbf{k}_{4}+\mathbf{q}\lambda_{1}}^{\dagger}c_{\mathbf{k}_{3}\lambda_{3}}\left\langle c_{\mathbf{k}_{3}-\mathbf{q}\lambda_{2}}^{\dagger}c_{\mathbf{k},v}\right\rangle ,
\end{align*}
where $\langle...\rangle$ refers to the expectation value taken over
the exciton ground state (fully occupied  valence band and fully empty conduction band).
Applying this technique to all the contributions, we find:
\begin{align*}
	& \left[\hat{H}_{{\rm int}},b_{\nu}^{\dagger}\right]=\\
	- & \sum_{\mathbf{k},\mathbf{p}}\frac{\phi_{\nu}(\mathbf{p})}{A^{3/2}}V_{\mathbf{k}-\mathbf{p}}F_{c,v,v,c}(\mathbf{k},\mathbf{p},\mathbf{k}-\mathbf{p})c_{\mathbf{k},c}^{\dagger}c_{\mathbf{k}v}\\
	- & \sum_{\mathbf{k},\mathbf{p}}\frac{\phi_{\nu}(\mathbf{p})}{A^{3/2}}V_{\mathbf{k}-\mathbf{p}}F_{v,c,v,c}(\mathbf{k},\mathbf{p},\mathbf{k}-\mathbf{p})c_{\mathbf{p},c}^{\dagger}c_{\mathbf{p},v}\\
	+ & \sum_{\mathbf{k},\mathbf{p}}\frac{\phi_{\nu}(\mathbf{k})}{A^{3/2}}V_{\mathbf{p}-\mathbf{k}}F_{v,v,v,v}(\mathbf{p},\mathbf{k},\mathbf{p}-\mathbf{k})c_{\mathbf{k},c}^{\dagger}c_{\mathbf{k},v}.
\end{align*}
Adding this expression with the result of the commutator with $\hat{H}_{0}$
and demanding the result to be the same as the one obtained using
the bosonic commutation relations we find

\begin{align}
	E_{\nu}\phi_{\nu}(\mathbf{k}) & =\phi_{\nu}(\mathbf{k})\left(E_{\mathbf{k},c}-E_{\mathbf{k},v}\right)\nonumber \\
	& -\frac{1}{A}\sum_{\mathbf{p}}\phi_{\nu}(\mathbf{p})V_{\mathbf{k}-\mathbf{p}}F_{c,v,v,c}(\mathbf{k},\mathbf{p},\mathbf{k}-\mathbf{p})\nonumber \\
	& +\frac{1}{A}\phi_{\nu}(\mathbf{k})\sum_{\mathbf{p}}V_{\mathbf{p}}\bigg\{ F_{v,v,v,v}(\mathbf{p}+\mathbf{k},\mathbf{k},\mathbf{p})\nonumber \\
	& -F_{v,c,v,c}(\mathbf{p}+\mathbf{k},\mathbf{k},\mathbf{p})\bigg\}.
\end{align}
The last term on the right hand side can be absorbed by $\phi_{\nu}(\mathbf{k})\left(E_{\mathbf{k},c}-E_{\mathbf{k},v}\right)$, and essentially describes a renormalization of the single-particle bands when interactions between electrons are accounted for, shifting the bands in energy. Neglecting this terms for the time being, we have
\begin{align}
	E_{\nu}\phi_{\nu}(\mathbf{k}) & =\phi_{\nu}(\mathbf{k})\left(E_{\mathbf{k},c}-E_{\mathbf{k},v}\right)\nonumber \\
	& -\frac{1}{A}\sum_{\mathbf{p}}\phi_{\nu}(\mathbf{p})V_{\mathbf{k}-\mathbf{p}}F_{c,v,v,c}(\mathbf{k},\mathbf{p},\mathbf{k}-\mathbf{p}).
	\label{eq:BSEtwoBands}
\end{align}
This is the BSE, and it determines the exciton energies $E_\nu$ and wave functions $\phi_\nu (\mathbf{k})$. It clearly resembling Eq. (\ref{eq:Wannier_Momentum}) apart from the term $F_{c,v,v,c}$.

To clearly extract the Wannier equation from the BSE, let us introduce some approximations. To start, we use the approximation
where the energy bands are parabolic near their bottom, i.e. $E_{\mathbf{k},\lambda}=E_{\mathbf{0},\lambda}+\hbar^{2}k^{2}/2m_{\lambda}$
(with $m_{\lambda}$ the effective mass on the band $\lambda$), and
write:
\begin{equation*}
	E_{\mathbf{k},c}-E_{\mathbf{k},v}\approx E_{g}+\frac{\hbar^{2}k^{2}}{2\mu_{eh}},
\end{equation*}
where $\mu_{eh}$ has been defined before as the reduced mass of the electron-hole
pair and $E_{g}$ is the energy difference between the top and bottom
of the valence and conduction bands, respectively. Then, the from
factor on the second term on the right hand side is approximated to
1, which is a good approximation 
for a large band gap. Using these three approximations, one finds:
\begin{align}
	E_{\textrm{bind.,}\nu}\phi_{\nu}(\mathbf{k}) & =\frac{\hbar^{2}k^{2}}{2\mu_{eh}}\phi_{\nu}(\mathbf{k})-\frac{1}{A}\sum_{\mathbf{p}}\phi_{\nu}(\mathbf{p})V_{\mathbf{k}-\mathbf{p}}.
\end{align}
with $E_{\textrm{bind.},\nu} = E_\nu  - E_g$. It is now a simple task to recognize a convolution integral on the last term
of the right hand side. Thus, Fourier transforming the whole equation
to real space, we obtain
\begin{equation}
	E_{\textrm{bind.,}\nu}\psi_{\nu}(\mathbf{r})=-\frac{\hbar^{2}\nabla^{2}}{2\mu_{eh}}\psi_{\nu}(\mathbf{r})-\psi_{\nu}(\mathbf{r})V(\mathbf{r}).
\end{equation}
which is the Wannier equation. 

Thus, we see that to obtain  the energies and wave functions of excitons in 2D systems two approaches can be followed. One either solves an integral equation in momentum space, or a differential equation in real space. The latter is easier to deal with, and more suitable for analytical approaches. The shortcomings of the real space approach are a general loss of accuracy due to the approximation used, and the loss of some physical phenomena associated with the form factor $F_{c,v,v,c}$ (in particular the real space approach fails to capture the lack of degeneracy between states with symmetric angular quantum numbers).

\subsection{Extension to multi-band systems}

In the derivation we just presented, it was assumed that we were working with a system with a single pair of bands. However, on may be interested in studying multi-band systems were more than one valence/conduction band should be considered.

For brevity, we shall only state the modified versions of the equations for the multi band case, and refer the reader to Refs. \cite{TGP2015_BSE,TGP2018_BSE} for more details.

For a multi band system, we define the state of an exciton as
\begin{equation}
	|\nu\rangle=\frac{1}{\sqrt{A}} \sum_{cv} \sum_{\mathbf{k}}\phi_{cv,\nu}(\mathbf{k})c_{\mathbf{k},c}^{\dagger}c_{\mathbf{k},v} |GS\rangle,
\end{equation}
which differs from the previous definition on the sum over the bands, which was previously absent.  The BSE is also modified, and reads
\begin{align}
	E_{\nu}\phi_{cv,\nu}(\mathbf{k})  &=\phi_{cv,\nu}(\mathbf{k})\left(E_{\mathbf{k},c}-E_{\mathbf{k},v}\right)\nonumber\\
	& -\frac{1}{A}\sum_{c',v',\mathbf{p}}\phi_{c'v'\nu}(\mathbf{p})V_{\mathbf{k}-\mathbf{p}}\notag \\ 
	&\times F_{c,v,v',c'}(\mathbf{k},\mathbf{p},\mathbf{k}-\mathbf{p}).
\end{align}
where once again we see the sum over the bands of the system.

Next, we shall derive the form of the electrostatic potential that should be used to solve either the BSE or the Wannier equation in 2D systems.

\subsection{The Rytova-Keldysh potential\label{sec:R-K}}

Two-dimensional semiconductors are characterized by a 2D polarization. This, together with the fact that the electrostatic field lines are for the most of it outside the semiconductor makes the electrostatic potential different from the Coulomb interaction between two charges in the bulk, as we now shall see  \cite{PhysRevB.84.085406}.

The  key to the derivation of the Rytova-Keldysh potential is the fact
that the charge fluctuations are proportional to the Laplacian of the
potential evaluated at the plane of the 2D material,  which is assumed to been surrounded by vacuum for simplicity \cite{PhysRevB.84.085406}.
Such a fact comes from the following considerations: the induced
charge density, $\delta n_{2D}\left(\mathbf{r}_{\parallel}\right)$, due to a point charge located at a distance $\mathbf{r}$ from the system is given by the
2D polarization in the usual way 
$
\delta n_{2D}\left(\mathbf{r}_{\parallel}\right)=-\nabla\cdot\mathbf{P}_{2D},
$
where the three-dimensional position vector is given by $\mathbf{r}=(\mathbf{r}_\parallel,z)$, and $\delta n_{2D}$ has units of charge per unit area.
The polarization itself is proportional to the total electric field
$
\mathbf{P}_{2D}=-\epsilon_0\chi_{2D}\nabla V\left(\mathbf{r}_{\parallel},z=0\right),
$
with $\chi_{2D}$ having dimensions of length. Therefore,

\begin{equation}
	\delta n_{2D}\left(\mathbf{r}_{\parallel}\right)=\epsilon_0\chi_{2D}\nabla^{2}V\left(\mathbf{r}_{\parallel},z=0\right).
\end{equation}
Let us write Poisson's equation as:
$
\nabla^{2}V\left(\mathbf{r}\right)=-e\left[n_{2D,+}+n\left(\mathbf{r}\right)\right]/ \epsilon_{0},
$
where $n_{2D,+}$ is the background positive charge density due to the atomic nuclei (in units of particles per area).
We now write the electronic density (in units of particles per area) as $n\left(\mathbf{r}\right)=-n_{2D,-}+\delta\left(\mathbf{r}\right)+\delta\left(z\right)\Delta \sigma\left(\mathbf{r}_{\parallel}\right)$,
where $n_{2D,-}$ is the neutralizing density of negative charge,
$\delta\left(\mathbf{r}\right)$ represents the density of a localized charge at position $\mathbf{r}$,
and $\delta\left(z\right)\Delta \sigma\left(\mathbf{r}_{\parallel}\right)$ is the induced charge
density fluctuation in the 2D material.   With these definitions Poisson's equation
reads

\begin{align}
	\nabla^{2}V\left(\mathbf{r}\right) 
	=-\frac{e}{\epsilon_{0}}\delta\left(\mathbf{r}\right)-\delta\left(z\right)\chi_{2D}\nabla^{2}V\left(\mathbf{r}_{\parallel},0\right),
\end{align}
where $e\Delta \sigma (\mathbf{r}_\parallel)=\delta n_{2D} (\mathbf{r}_\parallel)$.  Fourier transforming the previous
equation we obtain  [$\mathbf{k}=(\mathbf{k}_\parallel, k_z)$]
\begin{equation}
	-\left(k_{\parallel}^{2}+k_{z}^{2}\right)V\left(\mathbf{k}\right)=-\frac{e}{\epsilon_{0}}+k_{\parallel}^{2}\chi_{2D}V\left(\mathbf{k}_{\parallel},z=0\right).
\end{equation}
Solving for $V\left(\mathbf{k}\right)$ we find 
\begin{equation}
	V\left(\mathbf{k}\right)=\frac{e}{\epsilon_{0}}\frac{1}{k_{\parallel}^{2}+k_{z}^{2}}-\frac{k_{\parallel}^{2}}{k_{\parallel}^{2}+k_{z}^{2}}\chi_{2D}V\left(\mathbf{k}_{\parallel},z=0\right).
\end{equation}
Fourier transforming the previous equation in the $k_{z}$ coordinate to real space
(and taking $z=0$) we obtain
\begin{equation}
	V(\mathbf{k}_{\parallel})=\frac{e}{2\pi\epsilon_{0}}\frac{\pi}{k_{\parallel}}-\frac{\pi}{2\pi}k_{\parallel}\chi_{2D}V(\mathbf{k}_{\parallel}),
\end{equation}
where
\begin{equation}
	V\left(\mathbf{k}_{\parallel}\right)=\int_{-\infty}^{\infty}\frac{dk_{z}}{2\pi}V\left(\mathbf{k}_{\parallel},k_{z}\right).
\end{equation}
Solving for $V\left(\mathbf{k}_{\parallel}\right)$ we obtain
\begin{align}
	V\left(\mathbf{k}_{\parallel}\right) & = \frac{e}{2\epsilon_{0}k_{\parallel}}\frac{1}{1+k_{\parallel}/\kappa_{\parallel}},
\end{align}
where $1/\kappa_{\parallel}=\chi_{2D}/2$. The Fourier transform
of the potential to real space on the 2D material reads \cite{keldysh1979coulomb}
\begin{align}
	V\left(\mathbf{r}_{\parallel}\right) & =\frac{e}{2\epsilon_{0}}\int\frac{d\theta dk_{\parallel}}{(2\pi)^{2}}\frac{e^{i\mathbf{k}_{\parallel}\cdot\mathbf{r}_{\parallel}}\kappa_{\parallel}}{\kappa_{\parallel}+k_{\parallel}} \nonumber \\
	& = \frac{e}{4\pi\epsilon_{0}r_{0}}\frac{\pi}{2}\left[H_{0}\left(\frac{r}{r_{0}}\right)-Y_{0}\left(\frac{r}{r_{0}}\right)\right],
	\label{eq:Keldysh_potential}
\end{align}
where $r\equiv r_{\parallel}$ now stands for the in-plane radial position,  $r_{0}=1/\kappa_\parallel$, and $H_{0}\left(x\right)$ and $Y_{0}\left(x\right)$ are the Struve function and the Bessel function of the second kind, respectively.  The repulsive potential  representing the interaction among the electrons in a polarizable 2D semiconductor reads $V_{\mathrm{R-K}}(r)=eV\left(\mathbf{r}_{\parallel}\right)$.  Despite its complicated analytical form, this potential approaches the Coulomb potential at large distances, but diverges logarithmically near the origin.

In the following sections we will show different methods to solve the excitonic problem, both in real and momentum spaces, using the Rytova-Keldysh potential.

\section{Real space solution}
In this section we will present different methods to solve the excitonic problem in real space. For each method we shall focus on a concrete system, and give the results for the binding energies and wave functions of the associated excitons.  We start discussing tight-binding model of a hBN monolayer, from where we derive the eigenvalues and eigenfunctions used in the calculation of the optical conductivity using the BSE.

\subsection{Tight-binding model of hBN monolayer: from real space to momentum space \label{sec:tb_model_hBN}}

In this section we aim to briefly discuss the tight binding model
used to described the single particle electronic properties of gaped
Dirac systems. We shall focus on the case of hBN, and explain how
to extend the model to describe TMDs.

\begin{figure*}
.	\centering{}\includegraphics[scale=1]{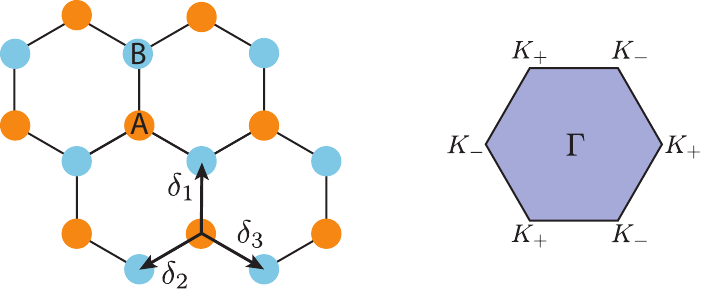}
	\caption{\label{fig:lattice}\textbf{Lattice and Brillouin zone of hBN.} (Left) Real space lattice of hBN, showing the non-equivalent $A$ and $B$ lattice sites,  and the lattice vectors connecting these two sites  (Right) Electronic Brillouin zone of hBN,  showing the $\Gamma-$point at the zone center and the non-equivalent $\mathbf{K}_+$ and $\mathbf{K}_-$ Dirac points. 
	}
\end{figure*}

The basic premise of a tight-binding model is that an electron in
a given orbital is located in the vicinity of its parent ion, but
may experience a transition, or hopping, to a an orbital of a nearby
ion. The likelihood of these transitions is characterized by the so
called \emph{hopping integrals}. In order for us to construct the
tight-binding Hamiltonian of hBN (or TMDs) we must first choose which
orbitals and which hoppings we wish to consider; obviously, increasing
the number of orbitals and the number of hoppings yields a more accurate
model, but also a more complex one.

n a minimal approach, we consider only the $p_{z}$ orbitals of the
atoms located on the two sublattices of the hBN layer, which we label
as $A$ and $B$ (see Fig.  \ref{fig:lattice}). Considering only hoppings between
nearest neighbors, we write the Hamiltonian in real space as
\begin{align*}
H & =\frac{E_{g}}{2}\sum_{i=1}^{N}|\boldsymbol{r}_{i},A\rangle\langle\boldsymbol{r}_{i},A|-\frac{E_{g}}{2}\sum_{i=1}^{N}|\boldsymbol{r}_{i},B\rangle\langle\boldsymbol{r}_{i},B|\\
 & +t\sum_{i=1}^{N}\sum_{j=1}^{3}|\boldsymbol{r}_{i},A\rangle\langle\boldsymbol{r}_{i}+\boldsymbol{\delta}_{j},B|+\textrm{h.c.}
\end{align*}
where $E_{g}/2$ is the on-site energy of the $p_{z}$ orbitals, $t$
is the hopping integral between an orbital at $\boldsymbol{r}_{i}$
(corresponding to the $A$ sublattice), and an orbital at $\boldsymbol{r}_{i}+\boldsymbol{\delta}_{j}$
(corresponding to the $B$ sublattice), where the $\boldsymbol{\delta}_{j}$
are the three vectors connecting a site of the $A$-sublattice to
its three nearest neighbors (all belonging to the $B$-sublattice).
Introducing the Fourier representation of $|\boldsymbol{r}_{i},A\rangle$,
that is, $|\boldsymbol{r}_{i},A\rangle=\frac{1}{\sqrt{N}}\sum_{\boldsymbol{p}}e^{i\boldsymbol{p}\cdot\boldsymbol{r}_{i}}|\boldsymbol{p},A\rangle$,
we are able to write the Hamiltonian as $H=\sum_{\boldsymbol{p}}\Psi_{\boldsymbol{p}}^{\dagger}H_{\boldsymbol{p}}\Psi_{\boldsymbol{p}}$,
with $\Psi_{\boldsymbol{p}}^{\dagger}=\left(|\boldsymbol{p},A\rangle,|\boldsymbol{p},B\rangle\right)$,
where the momentum space Hamiltonian reads
\begin{equation}
H_{\boldsymbol{p}}=\left(\begin{array}{cc}
E_{g}/2 & t\phi_{\boldsymbol{p}}\\
t\phi_{\boldsymbol{p}}^{*} & -E_{g}/2
\end{array}\right),
\end{equation}
with $\phi_{\boldsymbol{p}}=\sum_{j=1}^{3}e^{-i\boldsymbol{p}\cdot\boldsymbol{\delta}_{j}}$.
The energy bands and eigenvectors associated with this Hamiltonian
can be obtained through the diagonalization of $H_{\boldsymbol{k}}$.
The minimum gap between the valence and conduction bands occurs at
the Dirac points $K_{\pm}=\frac{2\pi}{a}\left(\pm\frac{2}{3\sqrt{3}},0\right)$, with $a$ the distance between nearest neighbours.
Since we will be interested on the low energy response alone, we can
expand $\phi_{\boldsymbol{k}}$ near these points, thus obtaining
an effective low energy Hamiltonian. Writing $\boldsymbol{p}=\tau\boldsymbol{K}+\boldsymbol{k}$,
with $\tau=\pm1$, and considering $|k|$ to be small, we find $\phi_{\boldsymbol{k}}\approx-\frac{3a}{2}\left(\tau k_{x}+ik_{y}\right)$.
Within this approximation, the effective Hamiltonian takes the form
of a Dirac Hamiltonian
\begin{equation}
H^{\textrm{eff}}=\frac{E_{g}}{2}\sigma_{z}-\hbar v_{F}\left(\tau k_{x}\sigma_{x}-k_{y}\sigma_{y}\right)\label{eq:H_eff}
\end{equation}
where $\hbar v_{F}=3ta/2$ is termed the Fermi velocity, and $\sigma_{i}$,
$i=\{x,y,z\}$ are the Pauli matrices. Diagonalizing $H^{\textrm{eff}}$,
we obtain the energy dispersion
\begin{equation}
E_{\pm,k}=\pm\sqrt{\frac{E_{g}^{2}}{4}+\hbar^{2}v_{F}^{2}k^{2}}
\label{eq:dispersion_hBN}
\end{equation}
where $+/-$ refer to the conduction/valence band. The eigenvectors
read
\begin{align}
|u_{+,\boldsymbol{k}}\rangle & =\left(\cos\frac{\xi_{k}}{2},-\sin\frac{\xi_{k}}{2}e^{-i\tau\theta}\right)^{\textrm{T}}\\
|u_{-,\boldsymbol{k}}\rangle & =\left(\sin\frac{\xi_{k}}{2}e^{i\tau\theta},\cos\frac{\xi_{k}}{2}\right)^{\textrm{T}}
\end{align}
where $\theta=\arctan k_{y}/k_{x}$ and $\xi_{k}=\arctan\left(2\hbar v_{F}k/E_{g}\right)$. We note in passing that where the complex exponentials $e^{\pm i\tau\theta}$ are located is important when the excitonic problem is treated in momentum space.

Near the bottom of the conduction band and near the top and of the valence band, the spectrum of hBN can be approximated by
Eq. (\ref{eq:dispersion_hBN}).
The effective mass follows as 
\begin{equation}
\frac{1}{m}=\left.  \frac{d^2 E_{\pm,k}}{\hbar^2dk^2}\right\vert_{k=0}=\pm\frac{2v_F^2}{E_g},\label{eq:mass}
\end{equation}
which leads to a reduced electron-hole mass  $\mu_{eh}=E_g/(4v_F^2)$.

Although the previous derivation focused on hBN, the broad strokes
remain the same to describe the single particle electronic structure
of TMDs. The two main differences for TMDs are the orbitals that must
be accounted for, and the necessity to account for spin-orbit coupling.
The effective Hamiltonian for such a system, can be obtained by adding
to Eq. (\ref{eq:H_eff}) the term \cite{Xiao2012,Benedikt2017}
\begin{equation}
H^{\textrm{SOC}}=\lambda\tau s_{z}\frac{\sigma_{z}-\mathbf{1}}{2}
\end{equation}
where $\mathbf{1}$ is the identity matrix, $s_{z}=\pm1$ labels the
spin projection of the bands and $\lambda$ quantifies the spin-orbit
coupling.

\subsubsection{Optical matrix element\label{subsec:matrix_element}}

An important quantity to compute, and that often appears when studying
the optical response, is the matrix element of the position operator.
We shall now find an analytical expression for that quantity. Consider
the matrix element
\begin{equation}
\langle u_{+,\boldsymbol{k}}|\boldsymbol{r}|u_{-,\boldsymbol{k}}\rangle.
\end{equation}
We now note that
\begin{equation}
\langle u_{+,\boldsymbol{k}}|\left[H,\boldsymbol{r}\right]|u_{-,\boldsymbol{k}}\rangle=\left(E_{+,k}-E_{-,k}\right)\langle u_{+,\boldsymbol{k}}|\boldsymbol{r}|u_{-,\boldsymbol{k}}\rangle
\end{equation}
Focusing on the commutator $\left[H,\boldsymbol{r}\right]$, we write
\begin{equation}
\left[H,\boldsymbol{r}\right]=i\hbar v_{F}\left(\tau\sigma_{x}\hat{\boldsymbol{x}}-\sigma_{y}\hat{\boldsymbol{y}}\right)
\end{equation}
We thus find [see Ref.  \cite{leppenen2020exciton} for the oscillator strength of excitons in Dirac materials]
\begin{align}
\langle u_{+,\boldsymbol{k}}|\boldsymbol{r}|u_{-,\boldsymbol{k}}\rangle & =i\hbar v_{F}\frac{\langle u_{+,\boldsymbol{k}}|\left(\tau\sigma_{x}\hat{\boldsymbol{x}}-\sigma_{y}\hat{\boldsymbol{y}}\right)|u_{-,\boldsymbol{k}}\rangle}{E_{+,k}-E_{-,k}}\nonumber \\
 & =\frac{i\hbar v_{F}}{E_{+,k}-E_{-,k}} \left(\cos^{2}\frac{\xi_{k}}{2}-e^{2i\tau\theta}\sin^{2}\frac{\xi_{k}}{2}\right)\left(\tau\hat{\boldsymbol{x}}+i\hat{\boldsymbol{y}}\right)
 \label{eq:dipolar_operaor}
\end{align}
From the above matrix element we see that only the $s$ and $d$
angular momenta couple to the electromagnetic field.

\subsection{Variational method \label{sec:variational_method}}

The variational method \cite{nesbet2002variational,epstein2012variation} is the simplest technique of choice for addressing the problem of finding the exciton  low-lying states,  corresponding to the least energetic excitons which appear further away from the conduction band edge. The method consists in proposing a trial wave function with unknown parameters, computing the expectation value of the Hamiltonian, $H$,  and minimizing it relatively to the unknowns. Formally, we define a function $\vert \psi_\alpha(\mathbf{r})\rangle$, where $\alpha$ represents the set of unknowns. The average $E_\alpha=\langle\psi_\alpha(\mathbf{r})\vert H\vert \psi_\alpha(\mathbf{r})\rangle$ is then computed and the $\alpha$'s are obtained from $\partial E_\alpha/\partial\alpha=0$.
Once the $\alpha$'s are known, this allow us to obtain the minimum value of $E_\alpha$.

This method has been used to describe excitons in black phosphorus \cite{Gomes_2021} and in TMDs \cite{MartinsQuintela2020,Gomes_2021,Henriques:21,PhysRevB.103.235402,PhysRevB.103.235412,PhysRevB.104.205433},  providing good results when bench marked against  the solution of the Bethe-Salpeter equation and the measured excitonic spectrum.
We note in passing that perturbative methods (not described in this work) has also been applied with success to the description of black phosphorus \cite{Joao_bp}. The exciton Stark shift electron absorption is another problem that can also be computed used variational wave functions in real space \cite{pedersen2016} as well as possibility of using fractional dimensions \cite{pedersen2016b}.

Let us consider excitons in a hBN monolayer and model the effective potential with the Rytova-Keldysh potential
\begin{equation}
	V_{\mathrm{R-K}}(r)=-\frac{e^2}{4\pi\epsilon_0 \epsilon_r}\frac{\pi}{2 r_{0}}\left[H_{0}\left(\frac{r}{r_{0}}\right)-Y_{0}\left(\frac{r}{r_{0}}\right)\right],\label{eq:RK_potential}
\end{equation}
where
$\epsilon_{r}$ is the average relative permittivity of the two media surrounding the monolayer, and the minus sign accounts for the attractive interaction between the electron and the hole in the exciton.

As Eq. (\ref{eq:RK_potential}) diverges as $ r\rightarrow 0 $, we consider variational wave functions inspired by the solution of the 2D Coulomb potential \cite{PhysRevA.43.1186}. The \emph{ansatz} for the ground state wave function reads:
\begin{equation}
	\Psi_{10}=C_{10} e^{-r / \beta_{10}},
	\label{eq:Psi10}
\end{equation}
where $ \beta_{10} $ is a variational parameter and $ C_{10} $ a normalization constant.  This represents an approximation to the excitonic ground state wave function  \cite{Schmitt_Rink_1985}. The wave functions for excited states can then be written as \cite{PhysRevA.43.1186,Castellanos_Gomez_2014,Gomes_2021}
\begin{align}
	\Psi_{20}&=C_{20}\left(1-d\,\frac{r}{\beta_{20}}\right) \mathrm{e}^{-r/ \beta_{20}} \nonumber\\
	\Psi_{2 x}&=C_{2 x} r\cos\theta \mathrm{e}^{-r/ \beta_{2x}} 
	\label{eq:Psi_excited}\\
	\Psi_{2 y}&=C_{2 y} r\sin\theta \mathrm{e}^{-r/ \beta_{2y}},\nonumber
\end{align}
where all three functions are orthogonal to each other by symmetry, $ C_{i} $ is a normalization constant, the  $ \beta $'s are variational parameters,  and  the $ d $ parameter is obtained imposing orthogonality between $ \Psi_{10} $ and $ \Psi_{20} $, and is given by 
\begin{equation}
	d=\frac{\beta_{10}+\beta_{20}}{2\beta_{10}}.
\end{equation}

Next we give the eigenenergies of the four variational states ($\Psi_{2 x}$ and $\Psi_{2 y}$ are degenerated) introduced above as well as a plot of the corresponding wave functions.

\begin{figure}
	\centering{}\includegraphics[scale=0.7]{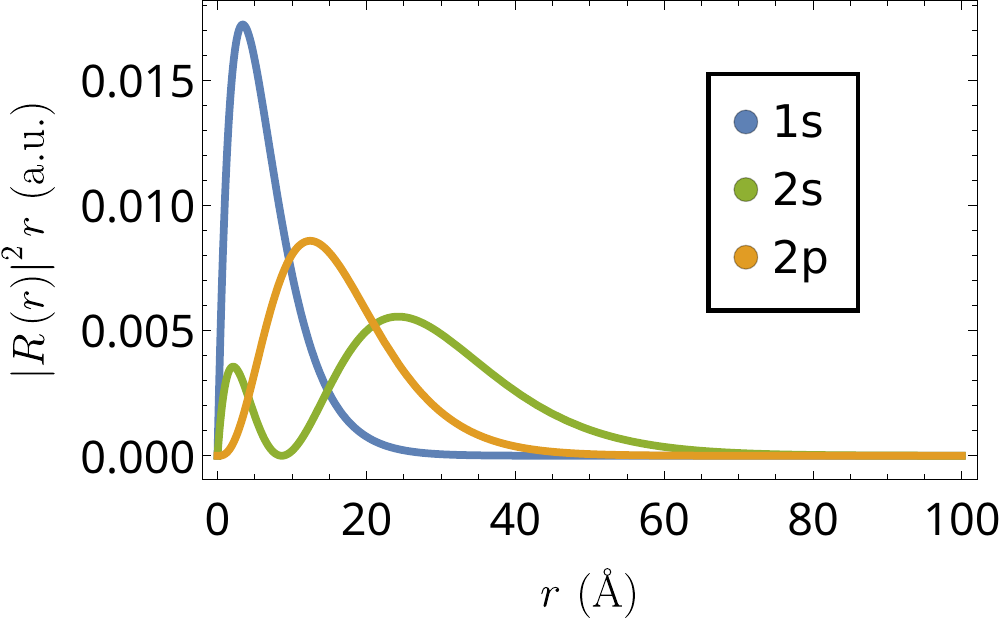}
	\caption{\label{fig:Variational_WF}\textbf{Variational  wave functions for hBN monolayer.} 
		Radial probability density $\left(\left|R(r)\right|^{2}r\right)$ of the $1s$, $ 2s $, and $ 2p $ states obtained using the variational method with the functions of Eq. (\ref{eq:Psi10}) and Eq. (\ref{eq:Psi_excited}) for hBN monolayer.  The radius is given in units of the Bohr radius $a_{0}$.}
\end{figure}

Focusing on the ground state for simplicity, we can then compute the expectation value of the Hamiltonian on the variational wave function as 
\begin{equation}
	E\left(\beta_{10}\right)=\matrixel{\Psi_{10}}{-\frac{\hbar^{2}\nabla^{2}}{2 \mu_{eh}} +V_{\mathrm{R-K}}(\mathbf{r})}{\Psi_{10}}.
\end{equation}
The previous average has analytical solution in terms of elementary functions.
Having computed this expectation value, we then find its minimum to determine the variational parameter $ \beta_{10} $. An analogous methodology is also applied to $ \Psi_{20} $ and $ \Psi_{2 x} $ (or $ \Psi_{2 y} $, as the two states are degenerate). The energies of the first three excitonic states in hBN are presented in Table \ref{tab:variational},  where the parameters used were $\epsilon_r=1$ and $r_{0}=10 $\AA  \cite{henriques_optical_2020}. The reduced mass $\mu_{eh}$ has been obtained directly from the hBN band structure near the Dirac points and it is given by Eq. (\ref{eq:mass}).
The excitonic variational wave functions of the $1s$, $ 2s $, and $ 2p $ states are depicted in Fig. \ref{fig:Variational_WF}.

\begin{table}
	\centering
	\begin{centering}
		\begin{tabular}{|c|c|c|c|}
			\hline 
			& \textbf{$\Psi_{10}$} & \textbf{$\Psi_{2x}$}& \textbf{$\Psi_{20}$} \tabularnewline
			\hline\hline 
			\textbf{$E_{n}^{\mathrm{var}}$} & $-1.27$ &  $-0.547$ & $-0.402$ 
			\tabularnewline
			\hline
			\textbf{$\beta_{n}$} & $6.796$ & $ 8.304 $& $ 11.84 $ 
			\tabularnewline
			\hline 
		\end{tabular}
		\par \end{centering}
	\caption{\label{tab:dielectric_comparison}
		\textbf{Variational energies for hBN monolayer. } Energies, in $ \mathrm{eV} $,  of the excitonic ground state and first excited states for hBN, together with the value of the corresponding $\beta_n$ variational parameter.	
	}
	\label{tab:variational}
\end{table}

\subsection{ Analytical expression for the optical sub-gap conductivity of hBN single layer\label{sec:conductivity_variational}}

In this section we take advantage of knowing the analytical form of the variational wave functions to obtain an analytical expression for the optical conductivity following the general definition introduced ahead [see Eq. (\ref{eq:opt_cond})].  Knowing the wave functions, however,  is not enough to determine the optical conductivity due to selection rules introduced by the matrix element of the dipolar operator. As a consequence some transitions are allowed (termed bright excitons) while others are not (termed dark excitons).  
A detailed study of the matrix element of the dipolar operator  reveals that the bright excitons correspond to the $s$ and $d$ transitions.  A study of the oscillator strength reveals that the dominant transition is the $1s$.  In the analytical expression given below we include the contribution from this  transition only,  as it dominates the sub-gap optical response of the monolayer.

For writing an analytical expression for the optical conductivity of
hBN, we need the Fourier transform to momentum space of the exciton
wave function in real space. We take advantage of the existence of
a simple variational wave function for the exciton.  Beginning with
the variational ansatz of Eq.  (\ref{eq:Psi10}),  we take the two-dimensional Fourier transform 
\begin{align}
\psi_{\mathrm{1s}}\left(k\right) & =2\pi\int_{0}^{\infty}\Psi_{10}\left(r\right){J}_{0}\left(kr\right)rdr\nonumber \\
 & =\frac{2\sqrt{2\pi}\beta_{10}}{\left(1+k^{2}\beta_{10}^{2}\right)^{3/2}},
\end{align}
where $J_0(x)$ is the cylindrical Bessel function of order zero.
As we previously noted,   Eq.  (\ref{eq:dipolar_operaor}),  only $s$ and $d$ angular momenta couple to
the electromagnetic field.  As the oscillator strength for
$d$-series transitions proves itself to be several orders of magnitude
smaller than that of the $s$-series states, 
we approximate the dipole matrix element  as
\[
\begin{aligned}\left.\left\langle u_{+,\mathbf{k}}|\mathbf{r}|u_{-,\mathbf{k}}\right\rangle \right|_{s} & \approx\frac{i\hbar v_{F}}{E_{+,k}-E_{-,k}}\cos^{2}\left(\frac{\xi_{k}}{2}\right)\left(\tau\hat{\boldsymbol{x}}+i\hat{\boldsymbol{y}}\right)\end{aligned}
.
\]
Focusing on $x$-polarized light, we write Eq. (\ref{eq:tg_pedersen_OMEGA}) as
\begin{align}
\sigma^{\left(1\right)}\left(\omega\right) & =\frac{e^{2}}{8\pi^{3}i\hbar}\frac{E_{1s}}{E_{1s}-(\hbar\omega+i\Gamma_{1s})}\left|\boldsymbol{\Omega}_{1s}\right|^{2}\\
 & =\frac{4e^{2}}{i\hbar}\frac{E_{1s}}{E_{1s}-(\hbar\omega+i\Gamma_{1s})}\left|\frac{\hbar^2 v^2_{F}-\hbar v_{F}E_{g}\beta_{10}}{E_{g}^{2}\beta_{10}^{2}-4\hbar^{2}v_{F}^{2}}+\frac{E_{g}\beta_{10}\hbar^2 v^2_{F}\,\mathrm{arcsec}\left(\frac{E_{g}\beta_{10}}{2\hbar v_{F}}\right)}{\left(E_{g}^{2}\beta_{10}^{2}-4\hbar^{2}v_{F}^{2}\right)^{3/2}}\right|^{2},
\end{align}
where the non-resonant term was excluded, $\Gamma_{1s}$ controls the line width of the excitonic absorption peak,    $\beta_{10}$ is the variational parameter 
entering in Eq.  (\ref{eq:Psi10})  (the $1s$ state),
and $E_{1s}=E_{bind.,1s}+E_{g}$ is the energy of the excitonic state. 
The optical conductivity above the gap requires the use of unbounded states \cite{leppenen2021sommerfeld}. 

\subsection{Expansion in a complete set of basis functions  \label{sec:complete_set}}

A common approach in physics for determining the spectrum and respective eigenfunctions of a system is to expand the unknown wave functions in a convenient complete set of basis functions, with unknown coefficients,  that is
\begin{equation}
	\psi_{\nu,l}(\mathbf{r})=\lim_{N\rightarrow \infty}\sum_{n=0}^{N} a_{\nu,n}\psi_{n, l}(\mathbf{r}),
\end{equation}
where $l$ is the angular quantum number (assuming a system with polar symmetry),  $\nu$ is the principal quantum number, $\psi_{n, l}(r)$ are the functions of the chosen basis,  and $a_{\nu,n}$ are the unknown coefficients.
The problem is then reduced to one in linear algebra, that is, to the determination of $a_{\nu,n}$. 

Here, opt to expand the exciton wave functions in terms of cylindrical Bessel functions, which form a complete basis on a disk of radius $R$. To obtain the eigenfunctions and eigenenergies of an infinite system, $R$ is chosen sufficiently large, such that the exponential tail of the wave functions is accurately captured. If this requirement is not satisfied, one is effectively solving the Wannier equation on a finite disk, which yields different solutions than those of the infinite (at least approximately) system. Just like in the variational methods, we focus on an hBN monolayer.

We begin considering the exact solution to the Schrödinger equation in polar coordinates for a particle confined by a circular infinite potential well. In this system, separation of variables can be immediately applied and a generic eigenstate reads \cite{Castano2005}
\begin{equation}
	\psi_{n, l}(r, \varphi)=\frac{C_{n, l}}{\sqrt{2 \pi}} e^{i l \varphi} J_{l}\left(z_{n, l} \frac{r}{R}\right),\label{eq:besselJ_function}
\end{equation}
where $J_{l}\left(x\right)$ is the Bessel function of the first kind
of order $l$, $z_{n,l}$ is the $n$-th zero of $J_{l}\left(z\right)$, and $C_{n,l}$ is a normalization constant given by 
\[
C_{n,l}=\sqrt{\frac{2}{R^{2}J_{\left|l\right|+1}^{2}\left(z_{n,l}\right)}}.
\]

As the Rytova-Keldysh potential, Eq. (\ref{eq:RK_potential}), is
invariant under rotations, the quantum number $l$ in Eq. (\ref{eq:besselJ_function}) is well-defined. 
As such, the Hamiltonian (Eq. (\ref{Sch_eq})) will be diagonal in blocks of fixed $l$, which significantly lowers the computational complexity, effectively reducing the integration to a 1D problem.  We can then compute the matrix elements of the Hamiltonian using two different functions $\psi_{n, l}(r, \varphi)$,
\begin{equation}
	H_{n,m}^{(l)}=\matrixel{\psi_{n, l}}{-\frac{\hbar^{2}}{2 \mu_{eh}} \nabla^{2}+V_{\mathrm{R-K}}(\mathbf{r})}{\psi_{m, l}}.
\end{equation}
With the matrix element  computed we write a matrix  for a chosen basis size $ N $.  Diagonalizing this $ N\times N $ matrix, we obtain an approximation of the eigenstates and eigenenergies of the Wannier equation with the Rytova-Keldysh potential. The eigenvectors of the matrix will give the coefficients $a_{\nu,n}$ of the expansion of the excitonic wave function in a basis of Bessel functions. 

\begin{figure}
	\centering{}\includegraphics[scale=0.7]{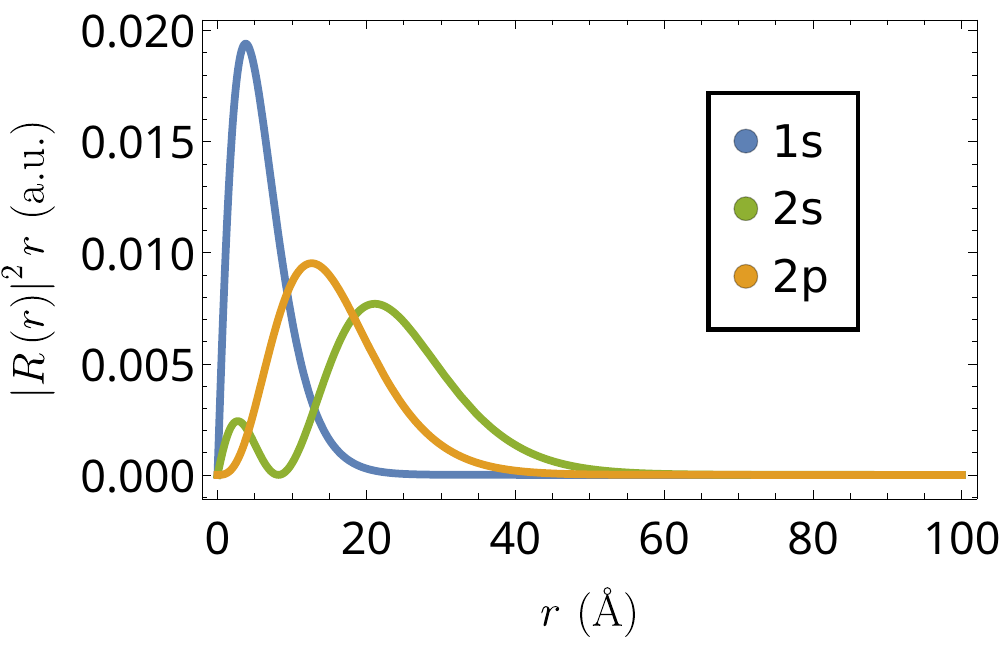}
	\caption{\label{fig:finite_basis}\textbf{Bessel function expansion wave functions for hBN monolayer.} 
		Radial probability density $\left(\left|R(r)\right|^{2}r\right)$ of the $1s$, $ 2s $, and $ 2p $ states obtained using the finite basis approximation for hBN monolayer.
		\label{fig:finite_basis_fig}}
\end{figure}

Focusing on a hBN monolayer, with the same parameters as before, we consider the $1s$, $ 2s $, and $ 2p $ excitonic states with a basis size of $ N=120$ and $R/a_0=200$.
\begin{table}
	\centering
	\begin{centering}
		\begin{tabular}{|c|c|c|c|c|}
			\hline 
			& \textbf{$1s$} & \textbf{$2s$} & \textbf{$2p_+$} & \textbf{$2p_-$} \tabularnewline
			\hline\hline 
			\textbf{$E_{n}$} & $-1.31$ & $-0.426$ & $-0.551$ & $-0.551$ 
			\tabularnewline
			\hline
			\textbf{$E^{\mathrm{BSE}}_{n}$} &  $-1.27$ & $-0.445$ & $-0.549$ & $-0.635$
			\tabularnewline
			\hline
			\textbf{$E^{\mathrm{var}}_{n}$} &  $-1.27$ & $-0.402$ &  $-0.547$ &  $-0.547$
			\tabularnewline
			\hline
		\end{tabular}
		\par \end{centering}
	\caption{
		\textbf{Energies from the basis-expansion method and from Bethe-Salpeter equation for hBN monolayer.   } Energies, in $ \mathrm{eV} $, of the excitonic ground state and first excited states for hBN obtained with the finite basis approach.	The last line of the table are the results of the variational calculation, which have a slightly larger values, as demanded by the variational theorem. Also note that the solution via the BSE predicts different energy values for \textbf{$2p_+$}  and \textbf{$2p_-$}, which is not predicted by the Wannier equation.
	}
	\label{tab:finite_basis}
\end{table}
The wave functions of the 1s,  2s, and 2p states are depicted in Fig. \ref{fig:finite_basis_fig} and the eigenenergies are 
given in Table \ref{tab:finite_basis}, both showing a very good agreement with the variational calculation.  Note how the wave functions already vanish for $r=100a_0$, justifying the use of a disk radius of $R=200a_0$. Increasing the value of $R$ would not produce any appreciable change on the final results. Reducing $R$, however, could significantly decrease the binding energies as a consequence of increased confinement.  Also, note that in Table
\ref{tab:finite_basis} we  give the result obtained via the solution of the Bethe-Salpeter equation (BSE) calculation (how to solve the BSE is described later in the text). This equation takes the spinorial structure 
of the wave function into account via the term
$F_{c,v,v,c}(\mathbf{k},\mathbf{p},\mathbf{k}-\mathbf{p})$, whereas
the Wannier equation makes the approximation 
$F_{c,v,v,c}(\mathbf{k},\mathbf{p},\mathbf{k}-\mathbf{p})=1$. Therefore, one should not except a complete agreement between the calculation using the BSE and using the expansion in the Bessel function's basis.  In any case, the energies found using  expansion in the Bessel function's basis are  smaller (more negative) than those found using the variational calculation, as expected.  Also note that 
including the spinorial structure into account leads to a degeneracy breaking between the states   \textbf{$2p_+$}  and \textbf{$2p_-$}.  

\subsection{Shooting method  \label{sec:shooting_method}}

The shooting method is a simple numerical approach for determining the eigenstates and the eigenvalues of the Wannier equation. In broad strokes, this method is based on the discretization of the Wannier equation, which is then integrated for different values of the exciton energy until a smooth wave function is obtained. Below we give a more detailed description of this procedure.

As we mentioned before, the Rytova-Keldysh potential has a logarithmic divergence at the origin. Because of this, to optimize the numerical caluclation, the differential equation should be discretized over a logarithmic grid, allowing for a sufficiently dense mesh near the origin (where the wave functions vary more abruptly  Hence, we first transform the radial Wannier equation using a logarithmic transformation and then apply an appropriate mesh of points for an accurate representation of the eigenstates over the full radial range. Here we address this method in the context of a TMD monolayer.

The basic premise of this method lies in the fact that the Wannier
equation  only yields continuous wave functions with a continuous derivative
for specific energy values, which correspond to the binding energies
of the excitonic states. Our task is thus to find what energy values
produce smooth wave functions. The shooting method is generally implemented
in the following manner: i) first, an initial guess for the binding
energy is chosen; ii) using this energy the Wannier equation is integrated
from left to right, and from right to left, matching the two results
somewhere far way from the edges of the domain. This can be done explicitly
by discretizing the differential equation, or by using an appropriate
numerical package. Also, one can choose to match the wave functions
themselves, or their logarithmic derivative; iii) next, the wave functions are scaled in the matching point, forcing the continuity of the complete wave function (this step is not necessary if one works with the logarithmic derivatives); iv) afterwards, the
continuity of the derivative of the resulting function must be verified;
if it is a smooth function, then the initial guess for the binding
energy was correct, if not the process must be repeated with a new
guess for the binding energy until the continuity condition is satisfied.

\begin{figure*}
	\centering{}\includegraphics[scale=0.9]{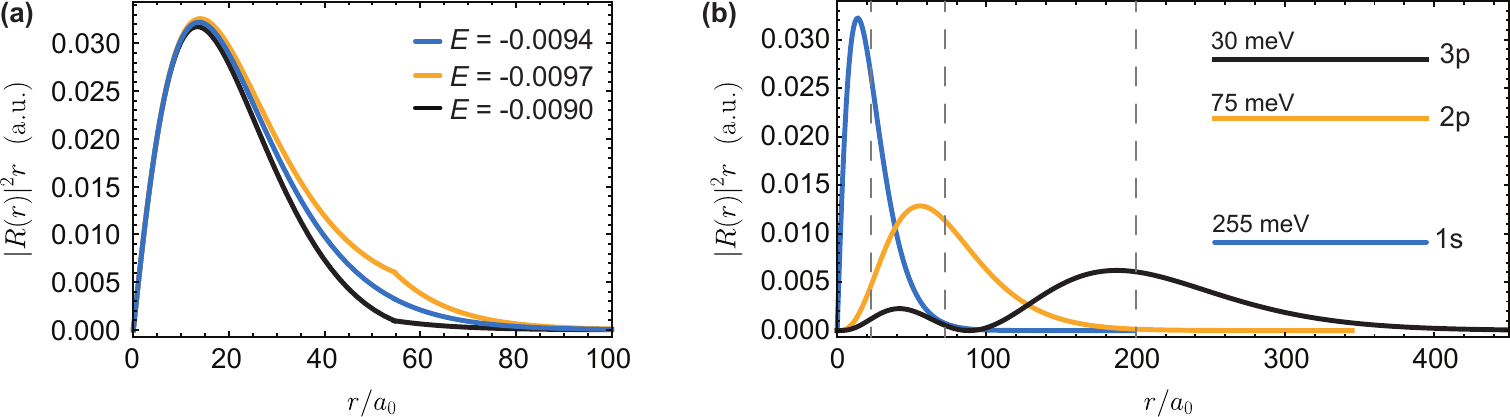}
	\caption{\label{fig:Shooting_WF}\textbf{Shooting method wave functions for WSe$_2$.} (a)
		Wave function of the $1s$ state of WSe$_2$ on a diamond substrate using three different energy guesses
		(whose values are presented in atomic units). The correct binding
		energy is $E=-0.0094$ since it leads to a smooth wave function. (b)
		Radial probability density $(|R(r)|^{2}r)$ of the the $1s$ state
		and the first excited states with $|m|=1$, obtained using the shooting
		method. The dashed lines mark the exciton Bohr radius of the three
		considered states. In both panels the radius is given in units of
		the Bohr radius $a_{0}$. The parameters used were 
		$\epsilon_r=3.32$ and $r_{0}=52 a_0$,
		taken from \cite{poellmann_resonant_2015},  with $a_0$ the Bohr radius; the reduced mass $\mu_{eh}=0.167 m_0$
		was obtained from the data presented in \cite{Kormanyos2015},  with $m_0$ the bare electron mass.
	}
\end{figure*}

In a more detailed way, we start with the Wannier equation, and write the wave function as $\psi(\mathbf{r})=R(r)\Phi(\theta)$,
with $\Phi(\theta)\propto e^{im\theta}$ where $m=0,\pm1,\pm2,...$
is the angular quantum number. Using this, the Wannier equation is
transformed into a differential equation defining the radial function
$R(r)$. Performing the substitution $u(r)=R(r)\sqrt{r}$, one finds
\begin{equation}
	-\frac{1}{2\mu_{eh}}\frac{d^{2}u}{dr^{2}}+\left[\frac{4m^{2}-1}{
		8\mu_{eh} r^{2}}+V_{\textrm{R-K}}(r)\right]u(r)=Eu(r),\label{eq:Wannier_Radial_u}
\end{equation}
where the term proportional to $m^{2}$ is the well known centrifugal
potential, and the boundary conditions are $u(r=0)=0$ and $u(r\rightarrow\infty)=0$.
To facilitate numerical calculations it is convenient to  consider
the boundary conditions $u(r=\epsilon)=\epsilon$ and $u(r=R)=\epsilon$ instead,
where $\epsilon$ is a small, yet finite, quantity and $R$ is chosen
sufficiently deep into the region where $u(r)$ decays monotonically
to zero, such that $u(R)\sim0$. Working with $u(r)$ instead of $R(r)$
has the advantage of eliminating terms with a first derivative in
the radial coordinate. Note that transforming the problem from $R(r)$
to $u(r)$ is a natural modification since $|u(r)|^{2}=|R(r)|^{2}r$
corresponds to the radial probability density. The shooting algorithm
can be directly applied to Eq. (\ref{eq:Wannier_Radial_u}), however,
to improve the efficiency of the calculation we can introduce the
change of variable $x=\ln r$, and the new function $u(x)=f(x)e^{x/2}$,
leading to the differential equation
	\begin{equation}
		-\frac{1}{2\mu_{eh}}\frac{d^{2}}{dx^{2}}f(x)+\left[\frac{m^{2}}{2\mu_{eh}}+e^{2x}V_{\textrm{R-K}}\left(e^{x}\right)\right]f(x)=e^{2x}f(x)E.
	\end{equation}

The advantage of working with a logarithmic grid 
lies in the improved description of the region near $r=0$, which
is critical when the differential equation is integrated from left
to right. Note that it is more important to have a detailed description
near $r=0$ than in the region $r\rightarrow\infty$, since in the
former the wave function and its derivative vary abruptly, while
in the latter the changes are more gradual due to the slow decaying
behaviour of the wave functions at large distances. Numerically, we define a minimum value of $x$ as $x_\mathrm{min}=e^{x_m}$, with $x_m$ a negative number and a maximum value of $x$ as $x_\mathrm{max}=e^{x_M}$, where $x_M$ is a positive number.  These two numbers,  $x_\mathrm{min}$ and $x_\mathrm{max}$, define the lower and upper limits of the spatial grid, respectively.
As noted,  $x_\mathrm{max}$ is chosen such that the wave functions are essentially zero in an appreciable position range in the grid of points.  The value of $x_\mathrm{min}$ is chosen such that the wave function near the origin is well represented.

In Fig. \ref{fig:Shooting_WF}a we depict the wave functions obtained
with the shooting method for the $1s$ state ($m=0$) of the TMD WSe$_2$,  using three
distinct guesses for the binding energy. Comparing the three results
it is clear that $E=-0.0094$ a.u. (255 meV) corresponds to the correct
binding energy, since it produces a smooth wave function, as opposed
to the other two guesses, whose wave functions present a sharp kink
in the matching point (implying that the used energy is not physically
meaningful). The binding energy of other $s-$states can be obtained
by changing the initial guess for the binding energy. To obtain the
solutions of other exciton states, e.g. the $2p$ state, we modify
the centrifugal potential (which now becomes finite), and introduce
new guesses for the binding energy. The plot of the radial probability
density of the $1s$, $2p$ and $3p$ states is depicted in Fig. \ref{fig:Shooting_WF}b.
There, we observe that the wave functions follow the same qualitative
traits as their counterpart for the 2D Hydrogen atom, with the difference
of being more spread out in space as a consequence of the reduced
interaction strength ($\ln r$ in 2D instead of $1/r$ in 3D,  when $r\rightarrow 0$). The dashed lines mark the exciton Bohr radius
of each state, corresponding to $23a_{0}$, $72a_{0}$ and $200a_{0}$,
with $a_{0}=0.53$\r{A}  the Bohr radius of the Hydrogen atom, for the $1s$, $2p$ and $3p$ states,
respectively.

\section{Momentum space solution \label{sec:bethe_salpeter}}

In the previous section we described three possible approaches to solve the excitonic problem in real space, i.e. to solve the Wannier equation. Now, in the current section, we will present a method to solve the excitonic problem directly in momentum space, i.e. to solve the BSE. As we mentioned when both equations were introduced, the BSE should be more accurate than then Wannier equation, since the latter is obtained from the former after the introduction of a series of approximations. This increased accuracy comes at the cost of a reduced number of approaches to solve the BSE (for a semi-analytical approach see \cite{PhysRevB.89.125309,GunnarPhD}).    

In what follows, we shall describe how to treat the BSE by first converting it from a 2D to a 1D integral equation, followed by its discretization and numerical diagonalization. To be specific, we shall consider the case of a biased bilayer graphene. The energies and wave functions of its excitons will be computed, and its optical conductivity determined.

\subsection{Tight binding model of Biased bilayer graphene}

Before describing how to solve the BSE for a multiband system, we shall briefly go over a tight binding model for biased bilayer graphene. This has been treated in other references where more detailed discussions can be found \cite{Castro_2010,PhysRevLett.99.216802,McCann2012,PhysRevB.74.161404}.

\begin{figure}[h]
	\centering
	\includegraphics[scale=0.7]{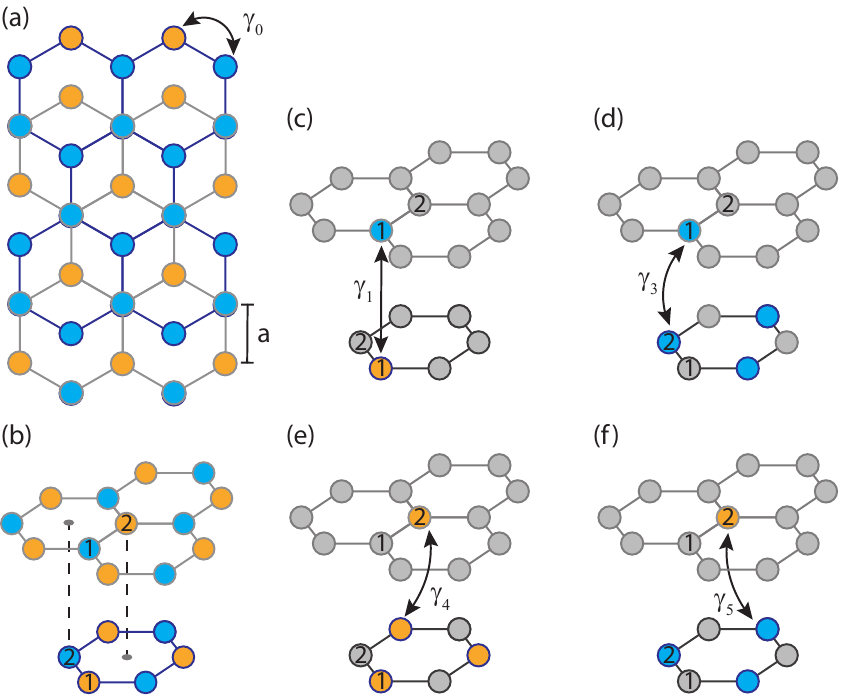}
	\caption{\textbf{Lattice structure of  graphene bilayer. }
	(Left) the Bernal stacking of graphene bilayer. (Right) hopping parameters defining the tight-binding model of  graphene bilayer
	}\label{fig:lattice_bilayer}
\end{figure}

In Fig.   \ref{fig:lattice_bilayer} we show the lattice of biased bilayer graphene. It essentially corresponds to two graphene layers in the AB (or Bernal) tacking configuration subject to an external electric field (which creates a potential difference between the bottom and top layers). To characterize this system we shall employ a tight binding Hamiltonian
written directly in momentum space, where we 
account for nearest neighbors intralayer hoppings ($\gamma_{0}$),
as well as nearest neighbors $(\gamma_{1})$ and next nearest neighbors
($\gamma_{3}$, $\gamma_{4}$ and $\gamma_{5}$) interlayer hoppings.
Doing so, one finds the Hamiltonian to be:
\begin{equation}
H_{\textrm{TB}}=\left[\begin{array}{cccc}
V & \gamma_{0}\phi(\mathbf{k}) & \gamma_{1} & \gamma_{4}\phi^{*}(\mathbf{k})\\
\gamma_{0}\phi^{*}(\mathbf{k}) & V & \gamma_{3}\phi^{*}(\mathbf{k}) & \gamma_{5}\phi(\mathbf{k})\\
\gamma_{1} & \gamma_{3}\phi(\mathbf{k}) & -V & \gamma_{0}\phi^{*}(\mathbf{k})\\
\gamma_{4}\phi(\mathbf{k}) & \gamma_{5}\phi^{*}(\mathbf{k}) & \gamma_{0}\phi(\mathbf{k}) & -V
\end{array}\right],
\end{equation}
when written in the basis $ \left\{\ket{1,b},\ket{2,b},\ket{1,t},\ket{2,t}\right\} $, where  $ b/t $ denotes bottom and top layers, and $ 1, 2 $ denotes the different sub-lattices in each layer (see Fig. \ref{fig:lattice_bilayer}).
Also $\phi(\mathbf{k}) = \sum_{i=1}^3 e^{i\mathbf{k}\cdot \boldsymbol{\delta}_i}$, with $\boldsymbol{\delta}_i$ are the vectors connecting a given site with its three nearest neighbors. The top layer is at a potential of $+V$, and the bottom layer at $-V$, meaning that the total potential difference between the top and bottom layers is $ 2V $. Although we heave considered several hopping parameters, we shall set $\gamma_3=\gamma_4=\gamma_5=0$ when computing the band structure and eigenvectors, due to their low impact on those results. However, when computing the optical response later in the text, these hoppings will be included in the definition of the dipole matrix element to more accurately capture the optical selection rules.

\begin{figure}[h]
	\centering
	\includegraphics[scale=0.7]{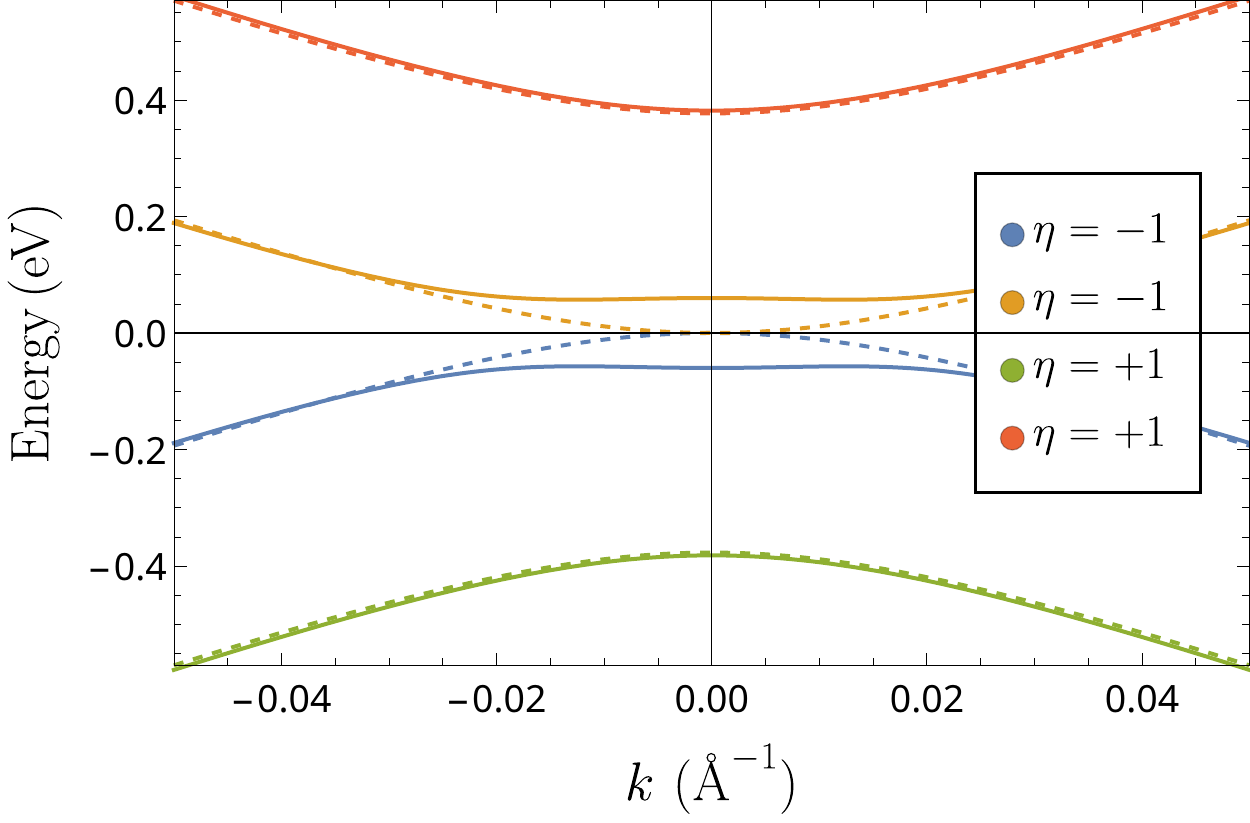}
	\caption{\textbf{Band structure of the biased graphene bilayer. }Electronic bands near the Dirac valley $\tau=1$ for a minimal model of biased Bernal--stacked
		bilayer graphene with bias potential $V=0\,\mathrm{meV}$ (dashed lines) and $V=60\,\mathrm{meV}$ (solid lines).}\label{fig:band_struct}
\end{figure}

Similar to that was done when the tight binding model for hBN was introduced, we can expand this Hamiltonian near the Dirac points $K_\pm = \left( \frac{4\pi}{3\sqrt{3}a}, 0 \right)$, leading to an effective low energy Hamiltonian. In Fig. \ref{fig:band_struct}, we depict the bands obtained from the diagonalization of such an Hamiltonian, for different values of $V$.
Clearly, when $V\neq0$, a gap opens; this is the key ingredient to explore excitonic effects.
From the diagonalization of the low energy Hamiltonian,
we also find the eigenvectors associated with each band. We label
these vectors as $|u_{\mathbf{k}}^{c,\eta}\rangle$ and $|u_{\mathbf{k}}^{v,\eta} \rangle$,
where the index $\eta$ is defined as in Fig. \ref{fig:band_struct},
i.e. the bands closer to the middle of the gap have $\eta=-1$, while
the other two have $\eta=1$. The eigenvectors read:
\begin{align}
|u_{\mathbf{k}}^{c,\eta=-1} \rangle & =\left[a_{c,1}^{-}e^{i\tau\theta},a_{c,2}^{-},a_{c,3}^{-}e^{i\tau\theta},a_{c,4}^{-}e^{2i\tau\theta}\right]\label{eq:low_spinors}\\
|u_{\mathbf{k}}^{c,\eta=+1} \rangle & =\left[a_{c,1}^{+},a_{c,2}^{+}e^{-i\tau\theta},a_{c,3}^{+},a_{c,4}^{+}e^{i\tau\theta}\right]\\
|u_{\mathbf{k}}^{v,\eta=-1} \rangle & =\left[a_{v,1}^{-}e^{-i\tau\theta},a_{v,2}^{-}e^{-2i\tau\theta},a_{v,3}^{-}e^{-i\tau\theta},a_{v,4}^{-}\right]\\
|u_{\mathbf{k}}^{v,\eta=+1} \rangle & =\left[a_{v,1}^{+},a_{v,2}^{+}e^{-i\tau\theta},a_{v,3}^{+},a_{v,4}^{+}e^{i\tau\theta}\right]
\end{align}
where $\theta=\arctan(k_y / k_x)$, with $\mathbf{k}=(k_x,k_y)$ the wave vector measured relatively to the Dirac point $\mathbf{K}_\tau$, where $\tau=\pm 1$. The specific form of each entry is of not particularly interesting for the current discussion due to their complicated analytical expressions. However, we note that for small $k$ one finds $|a_{c,1}^{-}|,|a_{c,3}^{-}|,|a_{c,4}^{-}|\ll|a_{c,2}^{-}|$,
$|a_{c,2}^{+}||a_{c,4}^{+}|\ll|a_{c,1}^{+}|,|a_{c,3}^{+}|$, $|a_{v,1}^{-}|,|a_{v,2}^{-}|,|a_{v,3}^{-}|\ll|a_{v,4}^{-}|$
and $|a_{v,2}^{+}||a_{v,4}^{+}|\ll|a_{v,1}^{+}|,|a_{v,3}^{+}|$.

\subsection{Solution of the BSE}

In this part of the section, we shall describe how to solve the BSE, in order to obtain the energies and wave functions of the excitons in biased bilayer graphene. As we saw before, the BSE for a multiband system can be expressed as \cite{PhysRevB.99.235433,PhysRevLett.120.087402,PhysRevB.104.115120}

	\begin{align}
		E \, \psi_{c,\eta_{1};v,\eta_{4}}\left(\mathbf{k}\right) & =\left(E_{\mathbf{k}}^{c,\eta_{1}}-E_{\mathbf{k}}^{v,\eta_{4}}\right)\psi_{c,\eta_{1};v,\eta_{4}}\left(\mathbf{k}\right)+\label{eq:BSE-simplified}\\
		&\quad+\sum_{\eta_{2},\eta_{3}}\sum_{\mathbf{q}}V\left(\mathbf{k}-\mathbf{q}\right)\left\langle u_{\mathbf{k}}^{c,\eta_{1}}\mid u_{\mathbf{q}}^{c,\eta_{2}}\right\rangle \left\langle u_{\mathbf{q}}^{v,\eta_{3}}\mid u_{\mathbf{k}}^{v,\eta_{4}}\right\rangle\psi_{c,\eta_{2};v,\eta_{3}}\left(\mathbf{q}\right) \nonumber
	\end{align}
where
$ \psi_{c,\eta_{1};v,\eta_{4}}\left(\mathbf{k}\right) $ is the excitonic wave function that we wish to obtain,  $\left|u_{\mathbf{k}}^{v/c,\eta}\right\rangle $ and $E_{\mathbf{k}}^{v/c,\eta}$
are the single particle electronic wave functions and energies of each band (which we obtained above), and $V\left(\mathbf{k}\right)$
is the Fourier transform of the Rytova-Keldysh potential coupling different bands and thus capturing
many-body effects including the intrinsic many-body nature of excitons.

As previously noted,  we consider the electrostatic potential to be the Rytova-Keldysh
potential \cite{rytova1967,keldysh1979coulomb} (usually employed to describe excitonic phenomena in mono- and few-layer
materials), whose derivation is available in Section \ref{sec:R-K}. In momentum space, this potential is given by 
\begin{equation}
	V\left(\mathbf{k}\right)=\frac{\hbar c\alpha}{\epsilon_r}\frac{1}{k\left(1+r_{0}k\right)},
\end{equation}
where $\alpha\approx1/137$ is the fine-structure constant, $\epsilon_r$
the mean dielectric constant of the medium above/below the bilayer graphene. The parameter $r_{0}$ corresponds to an in-plane screening length related
to the 2D polarizability of the material. It can be calculated from the single particle Hamiltonian of the system, although \emph{ab initio} calculations might be necessary for accurate computation of $ r_0 $ depending on the material\cite{acs.nanolett.9b02982}. This screening parameter varies with the bias potential $ V $ (since the band structure is also modified), and its numerical value is of the utmost importance if the excitonic properties of a specific system are to be studied accurately\cite{PhysRevB.92.245123,sponza2020proper}. An in-depth discussion of the in-plane screening length in bilayer graphene has been done in Ref. \cite{PhysRevB.99.035429}. In our biased bilayer graphene system, with $ V=52\,\mathrm{meV}$, the value of this effective screening length is $ r_{0}\approx 107.7\,\text{\AA} $.

To solve the Bethe-Salpeter equation, we first decompose the exciton wave function into the product of an angular and a radial part. Therefore, 
the wave function can be written as $\psi_{c,\eta_{1};v,\eta_{4}}\left(\mathbf{k}\right)=f_{c,\eta_{1};v,\eta_{4}}\left(k\right)e^{im\theta}$, with $m=0,\pm1,\pm2,...$  \cite{Chaves_2017}. Then, the phases of the eigenvectors $|u_{\mathbf{k}}^{c/v,\eta}$ must be chosen in such a way that the form factor $\left\langle u_{\mathbf{k}}^{c,\eta_{1}}\mid u_{\mathbf{q}}^{c,\eta_{2}}\right\rangle \left\langle u_{\mathbf{q}}^{v,\eta_{3}}\mid u_{\mathbf{k}}^{v,\eta_{4}}\right\rangle$ always leads to angular dependences of the form $e^{i \lambda (\theta_{\mathbf{k}} - \theta_{\mathbf{q}})}$, with $\lambda$ an integer, for every combination of valence and conduction bands. If this is the case, then the sum over $q$ can be converted into an integral, and through a change of variable the angular integral can be evaluated. This effectively transforms the initial 2D integral equation into a simpler 1D problem, which can then be numerically diagonalized. More details on how to achieve this are discussed in  \ref{app:BSE}.

Due to its particular band structure, the excitonic properties of biased bilayer graphene are dominated by the two bands closer to the middle of the gap, with the other two introducing only small corrections. As a result, to simplify the following analysis, we restrict our problem to those two bands. We note, however, that our description to solve the BSE could be equally applied if the 4 four bands were to be accounted for; in fact, in some systems, such as bilayer hBN, the complete four band model has to be used\cite{henriques2022limit}.

Based on these considerations, we solve the BSE, and obtain the binding energies displayed in Table  \ref{tab:BSE_eigenvalues}. The states were labelled following the notation of the Hydrogen atom, with s-states having $m=0$, p$_\pm$-states having $m=\pm1$, etc. We note the energy difference between states with symmetric $m$, something not captured by the Wannier equation, and that can be interpreted as a modification of the kinetic energy of the carriers due to the finite Berry curvature near the Dirac points in this type of material \cite{PhysRevLett.115.166803}.

Furthermore, it is important to note that Eq. (\ref{eq:BSE-simplified}) is actually a separate equation for each pair of bands $ c,\eta_{1};v,\eta_{4} $. This implies that there are $ 4 $ equations (2 valence times 2 conduction) that must be solved   simultaneously,  stemming from the two valence and two conduction bands.  
Additionally, as mentioned previously, a careful choice of the phases of the single-particle spinors allows us to transform the BSE into a $ 1D $ integral equation.  Details of this procedure are given in the Appendix of Ref.  \cite{PhysRevB.105.045411}. In \ref{app:BSE}, a summary overview of this methodology is discussed, as well as the numerical method for 1D integration of the BSE.

\begin{table}
	\centering
	\begin{centering}
		\begin{tabular}{|c|c|c|c|c|}
			\hline 
			& \textbf{$1s$} & \textbf{$2s$} & \textbf{$2p_+$} &  \textbf{$2p_-$}\tabularnewline
			\hline\hline 
			\textbf{$E_{n}^{\mathrm{BSE}}$} & $-16.3$ & $ -5.36 $ & $-9.35$ & $-10.3$
			\tabularnewline
			\hline
		\end{tabular}
		\par \end{centering}
	\caption{\textbf{Bethe-Salpeter equation eigenvalues for the  biased graphene bilayer.} Binding energies, in $ \mathrm{meV} $, of the excitonic ground state and first excited states for biased Bernal-stacked
		bilayer graphene encapsulated in hBN.	Note that the binding energies for $\textbf{$2p_+$}$ and 
		$\textbf{$2p_-$}$ are different a detail not captured by the solution of the Wannier equation.
	}
	\label{tab:BSE_eigenvalues}
\end{table}

\subsection{Optical conductivity}

Now, as the final part of this section, we shall use the solution of the BSE to compute the optical conductivity of biased bilayer graphene due to excitons.

Considering that the system is excited by an external time dependent electric field, and working in the dipole approximation, the optical conductivity of the system us given by 
 \cite{PhysRevB.92.235432, pedersen2021,sauer2021}  (generalizarions to nonlinear optics of 2D materials is also available \cite{TGP2015_BSE})
\begin{align}
\sigma^{(1)}(\omega)=\frac{e^{2}}{8\pi^{3}i\hbar} \sum_{n} E_n \frac{\boldsymbol{\Omega}_{n} \boldsymbol{\Omega}_{n}^{*}}{E_{n}-\hbar \omega}+\left(\omega\rightarrow-\omega\right)^{*},
\label{eq:opt_cond}
\end{align}
where the sum over $n$ represents the sum over excitonic states with energy $ E_n $ and wave function $ \psi_{n,cv} $. Additionally, a phenomenological broadening parameter $ \Gamma_n $ is included via the usual substitution $ \omega\rightarrow\omega+i\Gamma_n $. This parameter is considered to be $ n $-dependent \cite{PhysRevB.105.045411}.
In Eq. (\ref{eq:opt_cond}), $\boldsymbol{\Omega}_{n}$ is defined as
\begin{equation}
	\boldsymbol{\Omega}_{n}=\sum_{c,v}\sum_{\mathbf{k}}\psi_{n,cv}\left(\mathbf{k}\right)\left\langle u_{\mathbf{k}}^{v}\left|\mathbf{r} \cdot \hat{\boldsymbol{e}} \right|u_{\mathbf{k}}^{c}\right\rangle ,\label{eq:tg_pedersen_OMEGA}
\end{equation}
with $\left\langle u_{\mathbf{k}}^{v}\left|\mathbf{r}\right|u_{\mathbf{k}}^{c}\right\rangle$
the interband dipole operator matrix element; $\mathbf{r}$ is the position operator and $\hat{\boldsymbol{e}}$ the polarization vector of the external time dependent electric field. To compute this matrix element, we invoke the relation
\[
\left\langle u_{\mathbf{k}}^{v}\left|\mathbf{r}\right|u_{\mathbf{k}}^{c}\right\rangle =\frac{\left\langle u_{\mathbf{k}}^{v}\left|\left[H,\mathbf{r}\right]\right|u_{\mathbf{k}}^{c}\right\rangle }{E_{k}^{v}-E_{k}^{c}}.
\]
Evaluating the commutator and transforming the sum over $\mathbf{k}$ into a 2D integral, selection rules are imposed from the angular integration, i.e. the values of $m$ which lead to finite results are determined. As mentioned when the tight-binding model of this system was introduced, we consider the hoppings $\gamma_3$, $\gamma_4$ and $\gamma_5$ when evaluating this commutator. Although this approach is not entirely self consistent, it is an excellent approximation which significantly simplifies the calculations.

For a linearly polarized electric field, we reproduce the same optical selection rules as those found in \cite{PhysRevB.105.045411}, with the $ \gamma_{0} $ hopping parameter allowing transitions to states with $m=-\tau $ and $m=-3\tau $ (p- and f-states, respectively, when Hydrogenic labels are used). Transitions to states with $m=0$ (the s-states) are allowed only by the next nearest neighbors hopping parameter $ \gamma_{5} $, as obtained in \cite{PhysRevB.105.045411}. 

\begin{figure}
	\centering
	\includegraphics[scale=0.7]{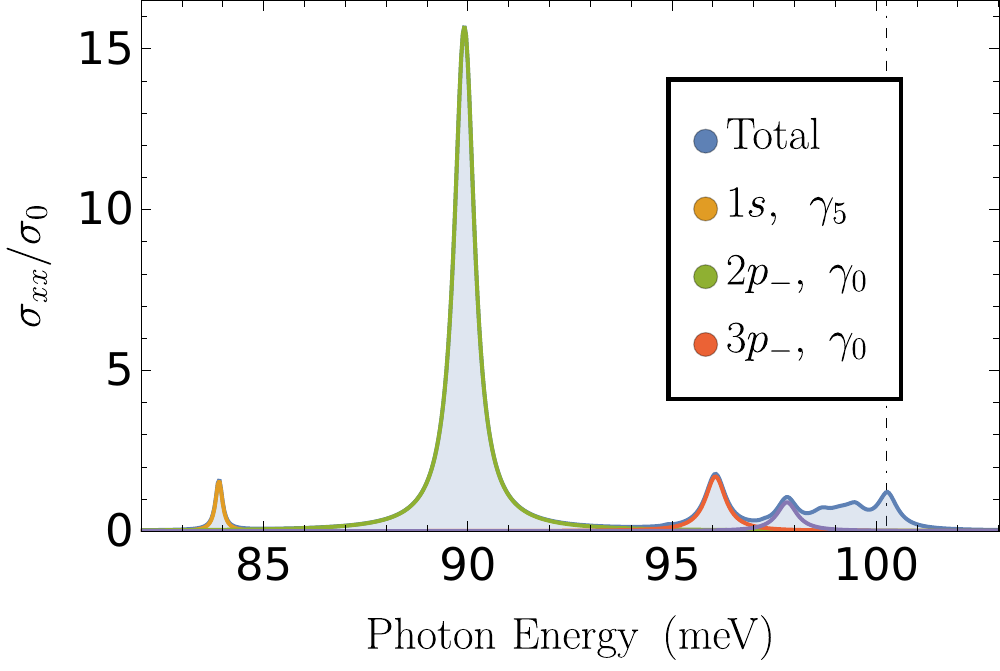}
	\caption{{\bf Optical conductivity of biased graphene bilayer.} Real part of the excitonic $xx$-conductivity for biased Bernal-stacked bilayer graphene encapsulated in hBN with a bias potential $V=52\,\mathrm{meV}$, broadening parameters $\Gamma_{np_-}=0.3\,\mathrm{meV}$ and $\Gamma_{ns}=0.1\,\mathrm{meV}$, and a $N=450$ point Gauss--Legendre quadrature. First ten states of each excitonic series were considered for the total conductivity. Vertical dashed lines represent the bandgap of the system. The different $ \gamma $s in the legend symbolize the hopping term that leads to specific resonances. The quantity $\sigma_0=\pi e^2/(2h)$ \label{fig:conductivity_xx_full}}
\end{figure}

Fixing the external bias potential at $V=52\,\mathrm{meV}$\cite{PhysRevB.105.045411}, we obtain Fig. \ref{fig:conductivity_xx_full} for the full $ xx $ linear optical conductivity. In this figure, we can clearly distinguish three resonances, namely those associated with $ 1s $, $ 2p_- $, and $ 3p_- $ states, with a plateau forming close to the bandgap value as the excitonic resonances become ever closer to each other.

\section{Conclusion}

In this tutorial we have addressed different theoretical methods for computing the excitonic wave function and corresponding eigenenergies. The methods vary.  Starting with the simpler variational method in real space and closing with the solution of the Bethe-Salpeter equation in momentum space, we hope to arm the reader with a multitude of techniques that allow their to start doing research on this topic. Putting together in a single paper the different methods scattered in the literature and making the text as self-contained as possible, the reader may find this reference quite useful for the daily research work.  On of the topics of this paper was the application of the BSE to biased graphene bilayer. This system can also be doped with a dual gate architecture, leading to the possibility of controlling 
excitons excitation by inducing residual carriers in the conduction band of the bilayer \cite{Knorr2022}.
The methods described in this paper can also be applied to more difficult systems, such as interlayer excitons in multi-layer materials or by considering spin-valley couplings. Additionally, a relevant system for future research is ABC--trilayer graphene, with an band gap controlled externally by an electric field perpendicular to the layers,  and which displays various many body phenomena recently considered in both experimental and theoretical papers. Also interesting to explore is the opening of a gap in twisted bilayer graphene due to Coulomb interactions \cite{APINYAN2020} and well as excitons in the metallic regime, a very unconventional situation \cite{Capua2021}. Finally, the extension of the methods discussed in this tutorial for studying excitons in Moiré TMD and hBN is worth pursuing.  

\section*{Acknowledgments}

M. F. C. M. Q. acknowledges the International Nanotechnology Laboratory (INL) and the Portuguese Foundation for Science and Technology (FCT) for the Quantum Portugal Initiative grant SFRH/BD/151114/2021.  J. C. G. H. acknowledges support by the Portuguese Foundation for Science and Technology (FCT) in the framework of the Strategic Funding UIDB/04650/2020.
N. M. R. P. acknowledges support by the Portuguese Foundation for Science and Technology (FCT) in the framework of the Strategic Funding UIDB/04650/2020, COMPETE 2020, PORTUGAL 2020, FEDER, and  FCT through projects POCI-01-0145-FEDER-028114, POCI-01-0145-FEDER-02888 and PTDC/NANOPT/ 29265/2017, PTDC/FIS-MAC/2045/2021, and from the European Commission through the project Graphene Driven Revolutions in ICT and Beyond (Ref. No. 881603, CORE 3).

\appendix
\section{Methodology for integration of Bethe-Salpeter Equation with Rytova-Keldysh potential: The case of biased graphene bilayer\label{app:BSE}}
Taking the thermodynamic limit, Eq. (\ref{eq:BSE-simplified}) can be written as
\begin{align}
	E \, f_{c,\eta_{1};v,\eta_{4}}\left(k\right) & =\left(E_{\mathbf{k}}^{c,\eta_{1}}-E_{\mathbf{k}}^{v,\eta_{4}}\right)f_{c,\eta_{1};v,\eta_{4}}\left(\mathbf{k}\right)-\label{eq:BSE-limit}\\
	&-\frac{1}{4\pi^{2}}\sum_{\eta_{2},\eta_{3}}\int qdqd\theta_{q}V\left(\mathbf{k}-\mathbf{q}\right)\left\langle u_{\mathbf{k}}^{c,\eta_{1}}\mid u_{\mathbf{q}}^{c,\eta_{2}}\right\rangle \left\langle u_{\mathbf{q}}^{v,\eta_{3}}\mid u_{\mathbf{k}}^{v,\eta_{4}}\right\rangle f_{c,\eta_{2};v,\eta_{3}}\left(q\right)e^{im\left(\theta_{q}-\theta_{k}\right)}.\nonumber
\end{align}
This problem can be simplified further, as $\left\langle u_{\mathbf{k}}^{c,\eta_{1}}\mid u_{\mathbf{q}}^{c,\eta_{2}}\right\rangle \left\langle u_{\mathbf{q}}^{v,\eta_{3}}\mid u_{\mathbf{k}}^{v,\eta_{4}}\right\rangle $
consists of a sum of different term with well--defined phases.   As such, it can be written as 
\begin{align}
	&	\left\langle u_{\mathbf{k}}^{c,\eta_{1}}\mid u_{\mathbf{q}}^{c,\eta_{2}}\right\rangle \left\langle u_{\mathbf{q}}^{v,\eta_{3}}\mid u_{\mathbf{k}}^{v,\eta_{4}}\right\rangle=\sum_{\lambda=0}^4\mathcal{A}_{\lambda}^{\eta_{1};\eta_{2};\eta_{3};\eta_{4}}\left(k,q\right)e^{i\lambda\left(\theta_{q}-\theta_{k}\right)},
\end{align}
where the angular dependence has been extracted from $\mathcal{A}_{\lambda}^{\eta_{1};\eta_{2};\eta_{3};\eta_{4}}\left(k,q\right) $.

Regarding the radial integral of the potential term, it can be written
as 
\begin{equation}
	I_{m}\left(k,q\right)=\int_{0}^{2\pi}\frac{\cos\left(m\theta\right)}{\kappa\left(k,q,\theta\right)\left[1+r_{0}\kappa\left(k,q,\theta\right)\right]}d\theta,
\end{equation}
where $\kappa\left(k,q,\theta\right)=\sqrt{k^{2}+q^{2}-2kq\cos\theta}$
and only the even term is non--zero due to parity. Inspecting the
integrand, it is clear that the $I_m$ function will be numerically
ill--behaved when $k=q$. For this effect, we decompose the integrand
in terms of partial functions as 
\begin{align}
	I_{m}\left(k,q\right) & =\int_{0}^{2\pi}\frac{\cos\left(m\theta\right)}{\kappa\left(k,q,\theta\right)}d\theta-r_{0}\int_{0}^{2\pi}\frac{\cos\left(m\theta\right)}{1+r_{0}\kappa\left(k,q,\theta\right)}d\theta\nonumber\\
	& =J_{m}\left(k,q\right)-K_{m}\left(k,q\right).\nonumber 
\end{align}
With this decomposition, it is clear now that only the $J_{m}\left(k,q\right)$
integral will be problematic when $k=q$. Substituting $I_{m}\left(k,q\right)$
into Eq. (\ref{eq:BSE-limit}), we write 
\begin{align}
	& 	E \, f_{c,\eta_{1};v,\eta_{4}}\left(k\right) = \left(E_{\mathbf{k}}^{c,\eta_{1}}-E_{\mathbf{k}}^{v,\eta_{4}}\right)f_{c,\eta_{1};v,\eta_{4}}\left(\mathbf{k}\right)-\nonumber \\
	& -\frac{1}{4\pi^{2}}\sum_{\eta_{2},\eta_{3}}\int_{0}^{+\infty}\sum_{\lambda=0}^4\left\{ J_{m+\lambda}\left(k,q\right)\mathcal{A}_{\lambda}^{\eta_{1};\eta_{2};\eta_{3};\eta_{4}}\left(k,q\right)-K_{m+\lambda}\left(k,q\right)\mathcal{A}_{\lambda}^{\eta_{1};\eta_{2};\eta_{3};\eta_{4}}\left(k,q\right)\right\} f_{c,\eta_{2};v,\eta_{3}}\left(q\right)qdq
\end{align}

Writing (where the possible values of $\lambda=0,\ldots 4$ is specific of the biased graphene bilayer)
\begin{align*}
	\mathcal{J}_{m}^{\eta_{1};\eta_{2};\eta_{3};\eta_{4}}\left(k,q\right)&=\sum_{\lambda=0}^4J_{m+\lambda}\left(k,q\right)\mathcal{A}_{\lambda}^{\eta_{1};\eta_{2};\eta_{3};\eta_{4}}\left(k,q\right),\\
	 \mathcal{K}_{m}^{\eta_{1};\eta_{2};\eta_{3};\eta_{4}}\left(k,q\right)&=\sum_{\lambda=0}^4K_{m+\lambda}\left(k,q\right)\mathcal{A}_{\lambda}^{\eta_{1};\eta_{2};\eta_{3};\eta_{4}}\left(k,q\right),
\end{align*}
the BSE can now be compactly written as 
\begin{align}
	& E \, f_{c,\eta_{1};v,\eta_{4}}\left(k\right) = \left(E_{\mathbf{k}}^{c,\eta_{1}}-E_{\mathbf{k}}^{v,\eta_{4}}\right)f_{c,\eta_{1};v,\eta_{4}}\left(\mathbf{k}\right)-\label{eq:final_BSE}\\
	&-\frac{1}{4\pi^{2}}\sum_{\eta_{2},\eta_{3}}\int_{0}^{+\infty}\left[ \mathcal{J}_{m}^{\eta_{1};\eta_{2};\eta_{3};\eta_{4}}\left(k,q\right)-\mathcal{K}_{m}^{\eta_{1};\eta_{2};\eta_{3};\eta_{4}}\left(k,q\right)\right] f_{c,\eta_{2};v,\eta_{3}}\left(q\right)qdq.\nonumber
\end{align}

We now focus our attention on the problematic $\mathcal{J}_{m}^{\eta_{1};\eta_{2};\eta_{3};\eta_{4}}\left(k,q\right)$
object. To treat the divergence at $q=k$, an auxiliary function $g_{m}\left(k,q\right)$
is introduced. This function obeys the limit 
\[
\lim_{q\rightarrow k}\left[\mathcal{J}_{m}^{\eta_{1};\eta_{2};\eta_{3};\eta_{4}}\left(k,q\right)-g_{m}\left(k,q\right)\right]=0
\]
and it modifies the integrals as
\begin{align}
	\int_{0}^{+\infty}\mathcal{J}_{m}^{\eta_{1};\eta_{2};\eta_{3};\eta_{4}}\left(k,q\right)f_{c,\eta_{2};v,\eta_{3}}\left(q\right)qdq\rightarrow & \int_{0}^{+\infty}\left[\mathcal{J}_{m}^{\eta_{1};\eta_{2};\eta_{3};\eta_{4}}\left(k,q\right)-g_{m}\left(k,q\right)\right]f_{c,\eta_{2};v,\eta_{3}}\left(q\right)qdq+\nonumber\\
	&\qquad\qquad+f_{c,\eta_{2};v,\eta_{3}}\left(k\right)\int_{0}^{+\infty}g_{m}\left(k,q\right)qdq.
\end{align}
Following \cite{PhysRevB.43.6530,portnoi,PhysRevB.105.045411}, this auxiliary function is chosen as 
\[
g_{m}\left(k,q\right)=\mathcal{J}_{m}^{\eta_{1};\eta_{2};\eta_{3};\eta_{4}}\left(k,q\right)\frac{2k^{2}}{k^{2}+q^{2}}.
\]

Having finished outlining the analytical procedure, we now proceed
to the numerical solution of the BSE. This is performed using the same methodology as \cite{PhysRevB.105.045411}, which we will quickly outline. 
A variable change is introduced
as to convert the integration limits from $\left[0,+\infty\right)$
to a finite limit, in this case $\left[0,1\right]$, defined as $q=\tan\left(\frac{\pi x}{2}\right)$. With this variable change, we proceed by discretizing $x$, writing the numeric
problem as 
\begin{align} 
	E \, f_{c,\eta_{1};v,\eta_{4}}\left(k\right) &= \left(E_{\mathbf{k}}^{c,\eta_{1}}-E_{\mathbf{k}}^{v,\eta_{4}}\right)f_{c,\eta_{1};v,\eta_{4}}\left(\mathbf{k}\right)+\frac{1}{4\pi^{2}}\sum_{\eta_{2},\eta_{3}}\sum_{j=1}^{N}\left[\mathcal{K}_{m}^{\eta_{1};\eta_{2};\eta_{3};\eta_{4}}\left(k_{i},q_{j}\right)f_{c,\eta_{2};v,\eta_{3}}\left(q_{j}\right)q_{j}\frac{dq}{dx_{j}}\right]\nonumber-\\
	& \qquad-\frac{1}{4\pi^{2}}\sum_{\eta_{2},\eta_{3}}\left\{ \sum_{j\neq i}\left[\mathcal{J}_{m}^{\eta_{1};\eta_{2};\eta_{3};\eta_{4}}\left(k_{i},q_{j}\right)f_{c,\eta_{2};v,\eta_{3}}\left(q_{j}\right)+g_{m}\left(k_{i},q_{j}\right)\right]q_{j}\frac{dq}{dx_{j}}w_{j}-\nonumber\right.\\
	&\qquad\qquad\left.-f_{c,\eta_{2};v,\eta_{3}}\left(k_{i}\right)\int_{0}^{\infty}g_{m}\left(k_{i},p\right)pdp\right\},\label{eq:BSE}
\end{align}
where $N$ is the number of points considered in the discretization, $w$ is the weight function of the quadrature in question, and the discretized variables are defined as $q_{i}\equiv q\left(x_{i}\right)$,
and $\frac{dq}{dx_{i}}\equiv\left.\frac{dq}{dx}\right|_{x=x_{i}}$.
It is important to note that, while $\int_{0}^{\infty}\mathcal{J}_{m}^{\eta_{1};\eta_{2};\eta_{3};\eta_{4}}\left(k,q\right)qdq$
is numerically problematic at $ q=k $, $\int_{0}^{\infty}g_{m}\left(k,q\right)qdq$
is well--behaved.

Regarding numerical integration, we specifically employ a Gauss--Legendre quadrature \cite{Kythe2002}, defined
as 
\[
\int_{a}^{b}f\left(x\right)dx\approx\sum_{i=1}^{N}f\left(x_{i}\right)w_{i},
\]
where 
\begin{align*}
	x_{i}&=\frac{a+b+\left(b-a\right)\xi_{i}}{2}, &w_{i}&=\frac{b-a}{\left(1-\xi_{i}^{2}\right)\left[\left.\frac{dP_{N}\left(x\right)}{dx}\right|_{x=\xi_{i}}\right]^{2}},
\end{align*}
with $\xi_{i}$ the $i$-th zero of the Legendre polynomial $P_{N}\left(x\right)$.

Finally, it is important to realize that Eq. (\ref{eq:final_BSE}) can be
written as the eigenvalue problem of a $tN\times tN$ matrix (\emph{i.e.},
a $t\times t$ matrix of $N\times N$ matrices). The $t^2$ blocks come from the different
combinations of band indices (\emph{i.e.}, $ t $ conduction bands times $ t $ valence bands), and each $N\times N$ matrix comes from the numerical discretization of the integral. Specifically, in biased bilayer graphene Eq. (\ref{eq:final_BSE}) would be written as a $ 2N\times 2N $ matrix.  Solving this eigenvalue problem for a sufficiently large quadrature, one obtains the excitonic eigenvalues and eigenfunctions.

\section*{Bibliography}

\bibliographystyle{unsrt}

\end{document}